\documentclass[aps,prd,superscriptaddress,10pt,showpacs,notitlepage,  
nofootinbib
]{revtex4-1}
\usepackage{bm}
\usepackage{amsmath,amssymb,amsthm}
\usepackage{latexsym,graphicx,color,subfigure}
\usepackage{enumerate}
\usepackage{amssymb}
\usepackage{hyperref}
\usepackage{mathtools}

\usepackage{graphicx}
\graphicspath{ {images/} }

\usepackage{color}

\newcommand{\be}{\begin{equation}}
\newcommand{\ee}{\end{equation}} 
\newcommand{\beq}{\begin{eqnarray}}
\newcommand{\eeq}{\end{eqnarray}}

\newcommand{\bea}{\begin{eqnarray}}
\newcommand{\eea}{\end{eqnarray}}

\def\tc{\textcolor{red}}
\def\tc2{\textcolor{blue}}
 
\renewcommand{\vec}[1]{\boldsymbol{#1}}

\allowdisplaybreaks[3]
\def\simge{\mathrel{
   \rlap{\raise 0.511ex \hbox{$>$}}{\lower 0.511ex \hbox{$\sim$}}}}
\def\simle{\mathrel{
   \rlap{\raise 0.511ex \hbox{$<$}}{\lower 0.511ex \hbox{$\sim$}}}}
\def\bigs{\mathrel{
   \rlap{\raise 0.531ex \hbox{$>$}}{\lower 0.531ex \hbox{$<$}}}}

\usepackage[normalem]{ulem}

\renewcommand\sout{\bgroup \color{blue} \ULdepth=-.5ex \ULset}

\begin{document}
\title{Phase structure of neutron $^{3}P_{2}$ superfluids in strong magnetic fields in neutron stars}
\author{Shigehiro Yasui}
\email{yasuis@keio.jp}
\affiliation{Department of Physics $\&$ Research and Education Center for Natural Sciences,\\ Keio University,Hiyoshi 4-1-1, Yokohama, Kanagawa 223-8521, Japan}
\author{Chandrasekhar Chatterjee}
\email{chandra@phys-h.keio.ac.jp}
\affiliation{Department of Physics $\&$ Research and Education Center for Natural Sciences,\\ Keio University,Hiyoshi 4-1-1, Yokohama, Kanagawa 223-8521, Japan}
\author{Muneto Nitta}
\email{nitta(at)phys-h.keio.ac.jp}
\affiliation{Department of Physics $\&$ Research and Education Center for Natural Sciences,\\ Keio University,Hiyoshi 4-1-1, Yokohama, Kanagawa 223-8521, Japan}
\date{\today}
\begin{abstract}
We discuss the effect of a strong magnetic field on neutron $^{3}P_{2}$ superfluidity.
Based on the attraction in the $^{3}P_{2}$ pair of two neutrons,
we derive the Ginzburg-Landau equation 
in the path-integral formalism
by adopting the bosonization technique and leave the next-to-leading order in the expansion of the magnetic field $B$.
We determine the $(T,B)$ phase diagram with temperature $T$, 
comprising three phases: 
the uniaxial nematic (UN) phase for $B=0$, 
D$_{2}$-biaxial nematic (BN) and 
D$_{4}$-BN phases 
in finite $B$ and strong $B$ such as magnetars, respectively,  
where  D$_{2}$ and D$_{4}$ are dihedral groups. 
We find that, 
compared with the leading order in the magnetic field known before,
the region of the D$_{2}$-BN phase in the $(T,B)$ plane is extended by the effect of the next-to-leading-order terms of the magnetic field.
We also present the thermodynamic properties, such as heat capacities and spin susceptibility, and find that the spin susceptibility 
exhibits anisotropies in the UN, D$_{2}$-BN, and D$_{4}$-BN phases.
This information will be useful to understand the internal structures of magnetars.
\end{abstract}
\maketitle


\section{Introduction}

Neutron stars are interesting astrophysical objects whose phenomena are induced by combinations of fundamental forces, i.e., strong interaction, weak interaction, electromagnetic interaction and gravitation (see Refs.~\cite{Graber:2016imq,Baym:2017whm} for recent reviews).
Studying neutron stars by several different signals can open a new approach to unveil the interiors in neutron stars.
It was epoch-making that gravitational waves from neutron star merger were observed directly~\cite{TheLIGOScientific:2017qsa}.
One of the most important properties of neutron stars is 
accompanied by a strong magnetic field.
It is known that the magnitude of the magnetic field is around $10^{12}$ G in standard neutron stars, and, in magnetars, it can reach about $10^{15}$ G at the surface and may reach even larger values
 in the inside~(see Refs.~\cite{Turolla:2015mwa,Kaspi2017} for reviews on magnetars).
As for the origin of the strong magnetic fields, researchers have studied several mechanisms such as spin-dependent interactions between neutrons~\cite{Brownell1969,RICE1969637,Silverstein:1969zz,Haensel:1996ss},\footnote{In the recent years, the many-body calculation leads to a negative result for the realization of strong magnetic fields~\cite{Bordbar:2008zz}.} the pion domain wall~\cite{Eto:2012qd,Hashimoto:2014sha}, the spin polarization in quark-matter core~\cite{Tatsumi:1999ab,Nakano:2003rd,Ohnishi:2006hs} and so on.
In neutron matter, the influence of the magnetic field on a neutron is provided by a finite magnetic moment, $\vec{\mu}_{n}=-(\gamma_{n}/2)\vec{\sigma}$ with the gyromagnetic ratio of a neutron, $\gamma_{n}=1.2 \times 10^{-13}$ MeV/T (in natural units, $\hbar=c=1$), and the Pauli matrices for the neutron spin $\vec{\sigma}$.
The interaction between the neutron and the magnetic field ($\vec{B}$)
supplies the energy splitting between the spin-up state and the spin-down state for a neutron, $-\vec{\mu}_{n}\!\cdot\!\vec{B}$.
For example, the strong magnetic field $|\vec{B}| \sim 10^{15}$ G in magnetars gives the mass splitting about ${\cal O}(10)$ keV (notice the unit relation, 1 T = $10^{4}$ G).

The dominant ingredient in neutron stars is the neutron matter, and it exists as superfluidity with a small admixture of superconducting protons and normal electrons (see Refs.~\cite{Chamel2017,Haskell:2017lkl} for recent reviews).
In relation to the observations of neutron stars, it is considered that 
the neutron superfluidity affects 
relaxation time after pulsar glitches~\cite{Reichely1969}, i.e., sudden speed-up events of 
neutron-star rotation.
It was proposed that the origin of pulsar glitches is the starquake in the core of the neutron stars~\cite{Baym1969,Pines1972,Takatsuka:1988kx} or the unpinning dynamics of neutron vortices pinned on the neutron-rich nuclei around the surface of the neutron stars~\cite{Anderson1975}. 
The neutron superfluidity is also related to rapid cooling by neutrino emissions known as the modified Urca process~\cite{Yakovlev:1999sk}.
The realization of a superfluid gap
can be supported by the recent observation of the rapid cooling time in neutron stars in Cassiopeia A~\cite{Heinke2010,Shternin2011,Page:2010aw}.
Interestingly, the nuclear forces can supply different types of pairings for two neutrons when the baryon number density in neutron matter changes from low density to high density~(see Ref.~\cite{Dean:2002zx} for a recent review).
In the early stage of the study of neutron superfluidity, Migdal considered the $^{1}S_{0}$ channel as the most attractive interaction~\cite{Migdal:1960}.
However, it was pointed out that the $^{1}S_{0}$ superfluidity cannot exist at higher densities due to the strong repulsion at short ranges of the nuclear force~\cite{1966ApJ...145..834W}.
Instead, the superfluidity is induced by the $^{3}P_{2}$ channel as the more attractive one in the higher-density region, which exists deep inside neutron stars.
The nuclear force for a pair of neutrons is provided by the $LS$ interaction at higher energy, and the gap equation for $^{3}P_{2}$ was analyzed by many researchers~\cite{Tabakin:1968zz,Hoffberg:1970vqj,Tamagaki1970,Takatsuka1971,Takatsuka1972,Takatsuka:1992ga,Baldo:1992kzz,Elgaroy:1996hp,Khodel:1998hn,Baldo:1998ca,Khodel:2000qw,Zverev:2003ak,Maurizio:2014qsa,Bogner:2009bt,Srinivas:2016kir}.
The angular momentum and the total spin are coupled to each other to form gap structures in the ground state, and the $\mathrm{U}(1)_{B} \times \mathrm{SO}(3)_{S} \times \mathrm{SO}(3)_{L} \times T \times P$ symmetry is broken into several subgroups ($B$ for baryon number, $S$ for spin rotation, $L$ for spatial rotation, $T$ for time-reversal symmetry, and $P$ for parity symmetry).
In nuclear physics, it is known that, in the $LS$ interaction, the $^{3}P_{2}$ channel is attractive, while the $^{3}P_{0}$ and $^{3}P_{1}$ channels are repulsive.
Thus, the $^{3}P_{2}$ pairing should be realized in neutron matter 
at high density.
It was discussed that the cooling process by neutrino emission can be described by low-energy excitations~\cite{Bedaque:2003wj,Bedaque:2012bs,Bedaque:2013fja,Bedaque:2014zta,Leinson:2009nu,Leinson:2010yf,Leinson:2010pk,Leinson:2010ru,Leinson:2011jr,Leinson:2012pn,Leinson:2013si,Leinson:2014cja} and by quantum vortices~\cite{Shahabasyan:2011zz}.

The neutron $^{3}P_{2}$ superfluidity has a rich gap structures due to the many possible combinations of spin and momentum~\cite{Fujita1972,Richardson:1972xn,Sauls:1978lna,Muzikar:1980as,Sauls:1982ie,Vulovic:1984kc,Masuda:2015jka,Masuda:2016vak}.
It is interesting to note that the $^{3}P_{2}$ pairings, i.e., tensor-type pairings, are analogous to those in $D$-wave superconductivity~\cite{Mermin:1974zz}.
It is known that $^{3}$He liquid below the critical temperature exhibits $P$-wave superfluidity in which many different states (BW state, ABM state, A$_{1}$ state, etc.) can be realized according to the values of temperature, pressure, magnetic field, and so on~\cite{vollhardt2013superfluid}.
Among them, for example, the BW state corresponds to the phase with the $^{3}P_{0}$ paring induced at low pressure and small magnetic field.
Chiral $P$-wave superconductivity has been established in Sr$_2$RuO$_4$~\cite{RevModPhys.75.657}. 
Apart from $P$-wave superfluidity, spin-2 superfluidity 
has been recently studied extensively as 
spin-2 Bose-Einstein condensates~(see Ref.~\cite{2010arXiv1001.2072K} 
for a review) that share some properties with  $^{3}P_{2}$ superfluidity.
Stimulated by those studies,
interesting topological properties in the neutron $^{3}P_{2}$ superfluidity were proposed:
topological superfluidity and 
gapless Majorana fermions on the boundary of $^{3}P_{2}$ superfluids~\cite{Mizushima:2016fbn}, 
a quantized vortex~\cite{Masuda:2015jka}, 
a soliton on it~\cite{Chatterjee:2016gpm} 
and  
a half-quantized non-Abelian vortex~\cite{Masuda:2016vak}.

The most fundamental equation for the superfluidity is the Bogoliubov--de-Gennes (BdG) equation.
In the BdG equation, the pairing gap function, which is dependent on position in general, can be solved self-consistently with the motion of the fermions.
The BCS equation, for which the BdG equation is restricted to 
the spatially uniform state of neutron $^{3}P_{2}$  superfluidity, 
was studied for a long time~\cite{Tabakin:1968zz,Hoffberg:1970vqj,Tamagaki1970,Takatsuka1971,Takatsuka:1992ga,Baldo:1992kzz,Elgaroy:1996hp,Khodel:1998hn,Baldo:1998ca,Khodel:2000qw,Zverev:2003ak,Maurizio:2014qsa,Bogner:2009bt,Srinivas:2016kir}.
More recently,
the BdG equation has been applied to investigate phase structures and topological properties in neutron $^{3}P_{2}$ superfluidity~\cite{Mizushima:2016fbn}.
Near the critical temperature, the BdG equation can be reduced to the Ginzburg-Landau (GL) equation as the low-energy effective theory.
In the GL equation, the pairing gap behaves as the effective degrees of freedom.
The GL equation was applied to study the phase structure of neutron $^{3}P_{2}$ superfluidity in a magnetic field~\cite{Fujita1972,Richardson:1972xn,Sauls:1978lna,Muzikar:1980as,Sauls:1982ie,Vulovic:1984kc,Masuda:2015jka,Masuda:2016vak}.
The ground state at the weak-coupling limit was found to 
be in a nematic phase~\cite{Sauls:1978lna},
where
 $\mathrm{U}(1)_{B} \times \mathrm{SO}(3)_{S} \times \mathrm{SO}(3)_{L} \times T \times P$ symmetry is broken into smaller (discrete) subgroups.
Either uniaxial nematic (UN) phase or 
biaxial nematic (BN) phase is realized,
depending on the magnitude of the magnetic field.
The UN phase is realized at zero magnetic field, 
while the BN phase is realized at finite magnetic field~\cite{Masuda:2015jka,Mizushima:2016fbn}.
Furthermore, the BN phase can be classified into two types: 
D$_{2}$-BN and D$_{4}$-BN phases,
characterized 
by the unbroken discrete (dihedral) symmetry, 
D$_{2}$ and D$_{4}$ symmetries, respectively. 
The GL equation was also applied to study the vortex structures of neutron $^{3}P_{2}$ superfluidity~\cite{Richardson:1972xn,Muzikar:1980as,Sauls:1982ie} and spontaneous magnetization in the vortices was found~\cite{Sauls:1982ie}.
Concerning the bulk properties in a magnetic field, 
it was obtained that the D$_{2}$-BN phase is realized at weak magnetic field, while the D$_{4}$-BN phase is realized at strong magnetic field 
(such as magnetars),
when only leading terms for the magnetic field in the GL equation were considered~\cite{Masuda:2015jka,Mizushima:2016fbn}.
It is not satisfactory, however, to conclude the influence of the strong magnetic field on neutron $^{3}P_{2}$ superfluidity without investigating effect by the next-to-leading-order terms in the strength of the magnetic field.
In fact, the neutron $^{3}P_{2}$ superfluidity inside neutron stars should be affected by strong magnetic fields. 
Because the typical magnitude of the neutron paring gap in the $^{3}P_{2}$ superfluidity is ${\cal O}(10^{2})$-${\cal O}(10^{3})$ keV, the neutron energy splitting by the strong magnetic field may change the properties of the neutron $^{3}P_{2}$ superfluidity. 
The purpose in the present paper is to clarify the effect of the strong magnetic field on the neutron $^{3}P_{2}$ superfluidity.
We specifically focus on the changes of the phase structures and the thermodynamical properties (heat capacity and spin susceptibility) by the strong magnetic field.

This paper is organized as the followings:
In Sec. \ref{sec:formalism}, we introduce the $LS$ interaction between two neutrons, and the spin interaction between the neutron and the magnetic field, and then derive the GL equation, including the next-to-leading-order terms of the magnetic field, from the BdG equation.
In Sec. \ref{sec:numerical_results}, we show our numerical results and clarify the effect on the phase structure by  the next-to-leading-order terms of the magnetic field.
We also present the thermodynamic properties (heat capacities and spin susceptibility), and discuss possible observables for neutron $^{3}P_{2}$ superfluidity inside neutron stars.
The final section is devoted to our conclusion.

\section{Formalism}
\label{sec:formalism}

\subsection{$LS$ potentials}

For the neutron field, we use a two-component spinor 
 $\varphi(t,\vec{x})$
in a nonrelativistic formalism.
The total Lagrangian is given by a sum of the kinematic term and the interaction term~\cite{Fujita1972,Richardson:1972xn,Sauls:1978lna,Muzikar:1980as,Sauls:1982ie,Vulovic:1984kc,Masuda:2015jka,Masuda:2016vak}: 
\begin{eqnarray}
  {\cal L}[\varphi]
=
   \varphi(t,\vec{x})^{\dag} \biggl( i\partial_{t} - \frac{\vec{\nabla}^{2}}{2m} - \mu + \vec{\mu}_{n} \!\cdot\! \vec{B} \biggr) \varphi(t,\vec{x})
+ G \sum_{a,b} T^{ab}(t,\vec{x})^{\dag}T^{ab}(t,\vec{x}),
\label{eq:action_tensor_0}
\end{eqnarray}
where $m=939$ MeV is a neutron mass and $\mu$ is the chemical potential of the neutron gas.
We have introduced the interaction term, $- \vec{\mu}_{n} \!\cdot\! \vec{B}$, for the magnetic moment of a neutron and the magnetic field $\vec{B} =(B_{1},B_{2},B_{3})$.
The second term 
is the interaction Lagrangian in the $^{3}P_{2}$ channel 
with the coupling constant $G>0$ as an attraction in the neutron $^{3}P_{2}$ pair.
$T^{ab}(t,\vec{x})$ ($a,b=1,2,3$; spin and space directions) 
is a symmetric and traceless tensor operator 
 defined by
\begin{eqnarray}
 T^{ab}(t,\vec{x})
= \frac{1}{2} \Bigl( \phi^{ab}(t,\vec{x}) + \phi^{ba}(t,\vec{x}) \Bigr) - \frac{1}{3} \delta^{ab} \sum_{c} \phi^{cc}(t,\vec{x}),
\label{eq:tensor_def}
\end{eqnarray}
where $\phi^{ab}(t,\vec{x})$ is the pairing function defined by
\begin{eqnarray}
    \phi^{ab}(t,\vec{x})
= - \varphi(t,\vec{x})^{t} \Sigma^{a\dag} \bigl( \nabla^{b}_{x} \varphi(t,\vec{x}) \bigr). 
\label{eq:phi_def}
\end{eqnarray}
Here, $\Sigma^{a} = i\sigma^{a}\sigma^{2}$ and $\nabla^{b}_{x}=\partial/\partial x^{b}$, 
where
$\sigma^{a}$ are the Pauli matrices, $\vec{\sigma}=(\sigma^{1},\sigma^{2},\sigma^{3})$, acting for neutron spin 
and
$\Sigma^{a}$ transforms as a vector for the rotation in space.
We consider only the tensor part in the $LS$ interaction, because the $^{3}P_{2}$ channel is only the attractive channel.
We notice that the interaction is of zero-range, 
and this simple form is sufficient for us in the present study.
In Eq.~(\ref{eq:phi_def}), the minus sign in the right-hand side is just 
for our convention.

\subsection{Derivation of GL free energy density}

We derive the GL equation from Eq.~(\ref{eq:action_tensor_0}) by introducing the gap function and integrate out the neutron fields (bosonization) based on the path-integral formalism.
The generating functional is written as
\begin{eqnarray}
  Z
=
  {\cal N}
  \int {\cal D}\varphi {\cal D}\varphi^{\dag}
  \exp
  \Biggl( i \int \mathrm{d}t \, \mathrm{d}\vec{x} \, {\cal L}[\varphi] \Biggr),
\end{eqnarray}
with ${\cal N}$ being an overall constant irrelevant to the dynamics.
The generating functional is transformed by introducing the auxiliary field, ${A}(t,\vec{x})$, which is a symmetric and traceless $3\times3$ tensor field with complex components.
Their components are denoted by ${A}(t,\vec{x})^{ab}$ in the basis of the spin and spatial directions ($a,b=1,2,3$).
By noting that multiplying
\begin{eqnarray}
 \int {\cal D}{A} {\cal D}{A}^{\ast}
 \exp\Biggl( i \int \mathrm{d}t \, \mathrm{d}^{3}\vec{x} \, \sum_{a,b} {A}(t,\vec{x})^{ab\ast} {A}(t,\vec{x})^{ab} \Biggr),
\end{eqnarray}
does not change the physical content of the generating functional,
we can express the generating functional $Z$ by 
\begin{eqnarray}
  Z'
&=&
  {\cal N}'
  \int {\cal D}\varphi {\cal D}\varphi^{\dag} {\cal D}{A} {\cal D}{A}^{\ast}
  \exp
  \Biggl( i \int \mathrm{d}t \, \mathrm{d}\vec{x} \, {\cal L}'[\varphi,{A}] \Biggr),
\label{eq:generating_functional_aux_3}
\end{eqnarray}
with an overall constant ${\cal N}'$,
where we have defined a new form of the Lagrangian by 
\begin{eqnarray}
   {\cal L}'[\varphi,{A}]
&=&
          \varphi(t,\vec{x})^{\dag} \biggl( i\partial_{t} - \frac{\vec{\nabla}^{2}}{2m} - \mu + \vec{\mu}_{n} \!\cdot\! \vec{B} \biggr) \varphi(t,\vec{x})
       + \sum_{a,b}
          \Bigl(
              {A}^{ab}(t,\vec{x})^{\ast} T^{ab}(\vec{x})
           + T^{ab}(\vec{x})^{\dag}  {A}^{ab}(t,\vec{x})
         \Bigr)
         \nonumber \\ && \hspace{0em}         
      + \frac{1}{G}
         \sum_{a,b} {A}^{ab}(t,\vec{x})^{\ast} {A}^{ab}(t,\vec{x}).
\end{eqnarray}
In the above derivation, ${A}(t,\vec{x})^{ab}$ has been shifted by a constant without changing the physical content.
${A}(t,\vec{x})^{ab}$ is the gap function as a matrix of spin ($a=1,2,3$) and angular momentum ($b=1,2,3$).
We assume the weak-coupling limit,\footnote{It was discussed that the strong-coupling effect does not change the phase diagram~\cite{Vulovic:1984kc}.} and consider the one-loop diagram for the neutron field.
Now performing the path integrals for $\varphi$, 
we obtain the effective action,
\begin{eqnarray}
 W'[{A}]
=
  - \mathrm{Tr} \ln \bigl( S_{m}(t,\vec{x})^{-1} \bigr)
  - \sum_{n\ge1} \frac{(-1)^{n+1}}{n} \mathrm{Tr} \bigl( S_{m}(t,\vec{x}) \hat{{A}}(t,\vec{x}) \bigr)^{n}
 + \frac{1}{4G}
    \mathrm{Tr}\bigl( {A}(t,\vec{x})^{\ast} {A}(t,\vec{x}) \bigr).
\label{eq:eff_pot_weakcouping_b}
\end{eqnarray}
Here we defined the neutron propagator,
\begin{eqnarray}
 S_{m}(t,\vec{x})^{-1}
&=&
   \left(
    \begin{array}{cc}
          i\partial_{t} - \frac{\vec{\nabla}^{2}}{2m} - \mu + \bar{\vec{\mu}}_{n} \!\cdot\! \vec{B}
           & 0 \\
          0 & i\partial_{t} + \frac{\vec{\nabla}^{2}}{2m} + \mu - \bar{\vec{\mu}}_{n}^{t} \!\cdot\! \vec{B}
    \end{array}
    \right),
\label{eq:Sm_propagator_def}
\end{eqnarray}
and the vertex function,
\begin{eqnarray}
   \hat{{A}}(t,\vec{x})
=
   \left(
    \begin{array}{cc}
          0 & - \sum_{a,b} {A}^{ab}(t,\vec{x}) t^{ab}(\vec{x})^{\dag} \\
          \sum_{a,b} {A}^{ab}(t,\vec{x})^{\ast} t^{ab}(\vec{x}) & 0
    \end{array}
    \right),
\label{eq:hat_tau_def}
\end{eqnarray}
with
\begin{eqnarray}
  t^{ab}(\vec{x})
&=&
- \biggl(
\frac{1}{2}\Sigma^{a\dag} \nabla^{b} + \frac{1}{2}\Sigma^{b\dag} \nabla^{a}
- \frac{1}{3}\delta^{ab} \sum_{c} \Sigma^{c\dag} \nabla^{c}
\biggr),
\label{eq:t_def_2}
\end{eqnarray}
which are expressed in the Nambu-Gor'kov formalism.
The trace ($\mathrm{Tr}$) is meant to take not only the sum over particle-hole and spin components but also the time-space integrals.
In Eq.~(\ref{eq:Sm_propagator_def}), we have defined $\bar{\vec{\mu}}_{n}=(\bar{\gamma}_{n}/2)\vec{\sigma}$ and $\bar{\gamma}_{n} = \gamma_{n}/(1+F_{0}^{a})$, where
$F_{0}^{a}$ is the Landau parameter in the Fermi-liquid theory 
introduced as a correction by the effect of the Hartree-Fock approximation.
This correction is necessary because the Hartree-Fock approximation is not covered in the present calculation for the particle-particle interaction at the one-loop level.

The effective potential (\ref{eq:eff_pot_weakcouping_b}) is regarded as the GL free-energy (GL equation).
Therefore, the GL free energy density is given by
\begin{eqnarray}
  f[{A}] = \frac{1}{2TV} W'[{A}],
\label{eq:eff_pot_weakcouping_b_f}
\end{eqnarray}
with $T$ and $V$ being the time and the volume of the system, $\displaystyle T=\int \mathrm{d}t\,1$ and $\displaystyle V=\int \mathrm{d}^{3}\vec{x}\,1$, respectively.
The factor $1/2$ in the right-hand side is necessary, because the effective potential (\ref{eq:eff_pot_weakcouping_b}) counts the doubled numbers of the degrees of freedom for neutrons in the Nambu-Gor'kov formalism.
Considering that ${A}(t,\vec{x})$ is a small quantity near the critical temperature and also that $\vec{B}$ is a small quantity, we expand Eq.~(\ref{eq:eff_pot_weakcouping_b_f}) with respect to ${A}(t,\vec{x})$ and $\vec{B}$ to  obtain the GL free-energy density:
\begin{eqnarray}
  f[{A}] = f_{0} + f_{6}^{(0)}[{A}] + f_{2}^{(\le4)}[{A}] + f_{4}^{(\le2)}[{A}] + {\cal O}(B^{m}{A}^{n})_{m+n\ge7}.
\label{eq:eff_pot_coefficient02_f}
\end{eqnarray}
We keep the terms up to 
${\cal O}({A}^{6})$, ${\cal O}(B^{2}{A}^{4})$, and ${\cal O}(B^{4}{A}^{2})$.
In the derivation, we have left only the lowest order for the derivative expansion for ${A}(t,\vec{x})$ by taking the long wavelength limit.
The first term $f_{0}$, which is irrelevant to the neutron paring, is given as
\begin{eqnarray}
f_{0}
=
  - T
    \int \frac{\mathrm{d}^{3}\vec{p}}{(2\pi)^{3}}
    \ln \Bigl(
             \bigl( 1+e^{-\xi_{\vec{p}}^{-}/T} \bigr)
             \bigl( 1+e^{-\xi_{\vec{p}}^{+}/T} \bigr)
         \Bigr),
\end{eqnarray}
with $\displaystyle \xi_{\vec{p}}^{\pm} = \xi_{\vec{p}} \pm |\vec{\mu}_{n}||\vec{B}|$ and $\displaystyle \xi_{\vec{p}}=\vec{p}^{2}/(2m)-\mu$.
The following terms are the interacting parts, in which the neutron parings are taken into account.
The momentum integrals and the Matsubara sums are estimated by assuming the particle-hole symmetry, i.e., the quasiclassical approximation, in the low-temperature region due to the weak coupling.
We use the notation $f_{n}^{(0)}$ for the free-energy density containing the ${A}$ field up to 
${\cal O}({A}^{n})$ at zero magnetic field, and $f^{(\le m)}_{n}[{A}]$ for the one containing the ${A}$ field up to 
${\cal O}({A}^{n})$ and the magnetic field up to 
${\cal O}(B^{m})$.
Their explicit forms read
\begin{eqnarray}
 f_{6}^{(0)}[{A}]
&=&
   \alpha^{(0)}
   \mathrm{tr}\bigl( {A}^{\ast} {A} \bigr)
\nonumber \\ &&
+ K^{(0)}
  \Bigl(
        \nabla_{xi} {A}^{ba\ast}
        \nabla_{xi} {A}^{ab}
     + \nabla_{xi} {A}^{ia\ast}
        \nabla_{xj} {A}^{aj}
     + \nabla_{xi} {A}^{ja\ast}
        \nabla_{xj} {A}^{ai}
  \Bigr)
\nonumber \\ &&
+ \beta^{(0)}
   \Bigl(
        \mathrm{tr}\bigl( {A}^{\ast} {A} \bigr) \mathrm{tr}\bigl( {A}^{\ast} {A} \bigr)
      - \mathrm{tr}\bigl( {A}^{\ast} {A}^{\ast} {A} {A} \bigr)
   \Bigr)
\nonumber \\ &&
+ \gamma^{(0)}
   \Bigl(
         - 3 \, \mathrm{tr}\bigl( {A} {A}^{\ast} \bigr) \, \mathrm{tr}\bigl( {A} {A} \bigr) \, \mathrm{tr}\bigl( {A}^{\ast} {A}^{\ast} \bigr)
        + 4 \, \mathrm{tr}\bigl( {A} {A}^{\ast} \bigr) \, \mathrm{tr}\bigl( {A} {A}^{\ast} \bigr) \, \mathrm{tr}\bigl( {A} {A}^{\ast} \bigr)
              \nonumber \\ && \hspace{3em} 
        + 6 \, \mathrm{tr}\bigl( {A}^{\ast} {A} \bigr) \, \mathrm{tr}\bigl( {A}^{\ast} {A}^{\ast} {A} {A} \bigr)
      + 12 \, \mathrm{tr}\bigl( {A}^{\ast} {A} \bigr) \, \mathrm{tr}\bigl( {A}^{\ast} {A} {A}^{\ast} {A} \bigr)
              \nonumber \\ && \hspace{3em} 
         - 6 \, \mathrm{tr}\bigl( {A}^{\ast} {A}^{\ast} \bigr) \, \mathrm{tr}\bigl( {A}^{\ast} {A} {A} {A} \bigr)
         - 6 \, \mathrm{tr}\bigl( {A} {A} \bigr) \, \mathrm{tr}\bigl( {A}^{\ast} {A}^{\ast} {A}^{\ast} {A} \bigr)
              \nonumber \\ && \hspace{3em} 
       - 12 \, \mathrm{tr}\bigl( {A}^{\ast} {A}^{\ast} {A}^{\ast} {A} {A} {A} \bigr)
      + 12 \, \mathrm{tr} \bigl( {A}^{\ast} {A}^{\ast} {A} {A} {A}^{\ast} {A} \bigr)
        + 8 \, \mathrm{tr}\bigl( {A}^{\ast} {A} {A}^{\ast} {A} {A}^{\ast} {A} \bigr)
   \Bigr),
\label{eq:eff_pot_w0_coefficient02_f}
\\
   f_{2}^{(\le4)}[{A}]
&=&
      \beta^{(2)}
      \vec{B}^{t} {A} {A}^{\ast} \vec{B}
+ \beta^{(4)}
   |\vec{B}|^{2}
   \vec{B}^{t} {A} {A}^{\ast} \vec{B},
\label{eq:eff_pot_B4w2_coefficient02_f}
\\
   f_{4}^{(\le2)}[{A}]
&=&
  \gamma^{(2)}
  \Bigl(
       - 2 \, |\vec{B}|^{2} \, \mathrm{tr}\bigl( {A} {A} \bigr) \, \mathrm{tr}\bigl( {A}^{\ast} {A}^{\ast} \bigr)
       - 4 \, |\vec{B}|^{2} \, \mathrm{tr}\bigl( {A} {A}^{\ast} \bigr) \, \mathrm{tr}\bigl( {A} {A}^{\ast} \bigr)
      + 4 \, |\vec{B}|^{2} \, \mathrm{tr}\bigl( {A} {A}^{\ast} {A} {A}^{\ast} \bigr)
      + 8 \, |\vec{B}|^{2} \, \mathrm{tr}\bigl( {A} {A} {A}^{\ast} {A}^{\ast} \bigr)
            \nonumber \\ && \hspace{2em}
        + \vec{B}^{t} {A} {A} \vec{B} \, \mathrm{tr}\bigl( {A}^{\ast} {A}^{\ast} \bigr)
       - 8 \, \vec{B}^{t} {A} {A}^{\ast} \vec{B} \, \mathrm{tr}\bigl( {A} {A}^{\ast} \bigr)
         + \vec{B}^{t} {A}^{\ast} {A}^{\ast} \vec{B} \, \mathrm{tr}\bigl( {A} {A} \bigr)
      + 2 \, \vec{B}^{t} {A} {A}^{\ast} {A}^{\ast} {A} \vec{B}
            \nonumber \\ && \hspace{2em}
      + 2 \, \vec{B}^{t} {A}^{\ast} {A} {A} {A}^{\ast} \vec{B}
       - 8 \, \vec{B}^{t} {A} {A}^{\ast} {A} {A}^{\ast} \vec{B}
       - 8 \, \vec{B}^{t} {A} {A} {A}^{\ast} {A}^{\ast} \vec{B}
  \Bigr).
\label{eq:eff_pot_B2w4_coefficient02_f}
\end{eqnarray}
The coefficients are given as
\begin{eqnarray}
   \alpha^{(0)}
&=&
   \frac{N(0)p_{F}^{2}}{3}
   \ln \frac{T}{T_{c0}},
\\
  K^{(0)}
&=&
   \frac{7 N(0) p_{F}^{4} \zeta(3)}{240m^{2}(\pi T)^{2}},
\\
  \beta^{(0)}
&=&
   \frac{7N(0)p_{F}^{4}\zeta(3)}{60\,(\pi T)^{2}},
\\
  \gamma^{(0)}
&=&
   - \frac{31N(0)p_{F}^{6}\zeta(5)}{13440\,(\pi T)^{4}},
\\
   \beta^{(2)}
&=&
    \frac{7\gamma_{n}^{2}N(0)p_{F}^{2}\zeta(3)}{48(1+F_{0}^{a})^{2}(\pi T)^{2}},
\\
   \beta^{(4)}
&=&
    - \frac{31\gamma_{n}^{4}N(0)p_{F}^{2}\zeta(5)}{768(1+F_{0}^{a})^{4}(\pi T)^{4}},
\\
  \gamma^{(2)}
&=&
  \frac{31\gamma_{n}^{2}N(0)p_{F}^{4}\zeta(5)}{3840(1+F_{0}^{a})^{2}(\pi T)^{4}},
\label{eq:eff_pot_coefficient0_parameters_FL_f}
\end{eqnarray}
with the condition that the temperature $T$ is close to the critical temperature at zero magnetic field $T_{c0}$; $|1-T/T_{c0}| \ll 1$.
$\zeta(n)$ is the zeta function.
The trace ($\mathrm{tr}$) is meant to be a sum over the spin and spatial directions.
The terms up to 
${\cal O}({A}^{6})$ and ${\cal O}(B^{2}{A}^{2})$ were known in the previous studies~\cite{Fujita1972,Richardson:1972xn,Sauls:1978lna,Muzikar:1980as,Sauls:1982ie,Vulovic:1984kc,Masuda:2015jka,Masuda:2016vak}.
We notice that the Fermi-surface approximation was adopted in the momentum integrals, and that the terms proportional to $|\vec{B}|^{2}\,\mathrm{tr}\bigl( {A} {A}^{\ast} \bigr)$ or to $|\vec{B}|^{4}\,\mathrm{tr}\bigl( {A} {A}^{\ast} \bigr)$ vanish in the present approximation. 
New terms found in the present study are the terms of ${\cal O}(B^{4}{A}^{2})$ in the second term in the right-hand side of Eq.~(\ref{eq:eff_pot_B4w2_coefficient02_f}), and  the terms of ${\cal O}(B^{2}{A}^{4})$ in Eq.~(\ref{eq:eff_pot_B2w4_coefficient02_f}).
Those new terms are important to investigate how the ground-state properties can be changed by strong magnetic fields.

The value of $T_{c0}$ is related to the coupling constant $G$ by
\begin{eqnarray}
 T_{c0}
=
 \frac{\pi e^{-\gamma}D}{8}
 \exp \biggl( -\frac{3}{8N(0)p_{F}^{2}G} \biggr),
\end{eqnarray}
with the Euler gamma $\gamma$, where the energy scale $D$ relevant for the dynamics is introduced in the momentum space below and above the Fermi surface.
We define the density of states at the Fermi surface, $N(0)=m\,p_{F}/(2\pi^{2})$ with the Fermi momentum $p_{F}=(3\pi^{2}n)^{1/3}$ with $n$ being the neutron number density. 
The chemical potential can be estimated as $\mu \approx p_{F}^{2}/(2m)$ by assuming small interactions as well as the low-temperature limit.

\section{Numerical results}
\label{sec:numerical_results}

\subsection{Phase diagrams}

We investigate the phase diagram of neutron $^{3}P_{2}$ superfluids based on the GL free energy (\ref{eq:eff_pot_coefficient02_f}).
In the numerical calculation, we use the parameter setting:
$T_{c0}=0.2$ [MeV],
$n=0.17$ [fm$^{-3}$] ($p_{F}=338$ [MeV]),
$F_{0}^{a}=-0.75$.
The Landau parameter $F_{0}^{a}$ is given by referring the value in $^{3}$He liquid at low temperature.

There are several phases in condensations with total spin two:  nematic (UN, D$_{2}$-BN, D$_{4}$-BN), cyclic and ferromagnetic phases~\cite{Mermin:1974zz,2010arXiv1001.2072K}. 
Among them, the nematic phase is realized at zero or weak magnetic field in the weak-coupling limit~\cite{Sauls:1978lna,Muzikar:1980as,Sauls:1982ie,Vulovic:1984kc,Masuda:2015jka,Masuda:2016vak}. 
In the nematic phase, the ${A}$ field can be parametrized by two variables ${A}_{0}$ and $r$:
\begin{eqnarray}
  {A}(t,\vec{x})
=
{A}_{0}
\left(
\begin{array}{ccc}
 r & 0  & 0  \\
 0 & -(1+r)  & 0  \\
 0 & 0 & 1  
\end{array}
\right),
\label{eq:tau_gs}
\end{eqnarray}
by restricting the region of the variables as ${A}_{0}\ge0$ and $-1\le r \le -1/2$, without loss of generality.
The values of ${A}_{0}$ and $r$ are constant numbers in time and space.
In each phase, ${A}$ has continuous or discrete unbroken symmetries: O(2) symmetry in the UN phase ($r=-1/2$), D$_{2}$ symmetry in the D$_{2}$-BN phase ($-1<r<-1/2$), and D$_{4}$ symmetry in the D$_{4}$-BN phase ($r=-1$)~\cite{Masuda:2015jka,Masuda:2016vak}. 
See Table~\ref{table-sym} for unbroken symmetries for each $r$ 
and the corresponding 
order-parameter manifolds and lower-dimensional homotopy groups.
For zero magnetic field, all states are degenerate in $r$ 
if the sixth order term in ${A}$ is neglected, 
in which case the symmetry of the potential is enhanced to $\mathrm{SO}(5)$ 
and the total order-parameter space is $[S^1 \times S^4 ]/{\mathbb Z}_2$~\cite{Uchino:2010pf}. 
Turning on the magnetic field or the sixth-order term in ${A}_{0}$, 
$r$ is determined by minimizing the free energy~\cite{Masuda:2015jka,Masuda:2016vak}. 
\begin{table*}[t!]
\begingroup
\renewcommand{\arraystretch}{1.5}
 \begin{tabular}{|c|c|c|c|ccccc|}
 \hline 
    $r$ & Phases & $H$ & $G/H$ &  $\pi_0$ &  $\pi_1$ & $\pi_2$ & $\pi_3$ & $\pi_4$ 
    \\ \hline \hline 
    $-1/2$ & UN & $\mathrm{O}(2)$ &$ \mathrm{U}(1) \times [\mathrm{SO}(3) / \mathrm{O}(2)]$ & $0$ & ${\mathbb Z} \oplus {\mathbb Z}_2$ & ${\mathbb Z}$ & ${\mathbb Z}$ & ${\mathbb Z}_2$ 
    \\ 
    $-1 < r < -1/2$ & ${\rm D}_2$-BN  & ${\rm D}_2$ & $ \mathrm{U}(1) \times [\mathrm{SO}(3) / {\rm D}_2]$ & $0$ &  ${\mathbb Z} \oplus {\mathbb Q}$ & $0$ & ${\mathbb Z}$ & ${\mathbb Z}_2$ 
    \\ 
    $-1$ & ${\rm D}_4$-BN & ${\rm D}_4$ & $ [\mathrm{U}(1) \times \mathrm{SO}(3)] / {\rm D}_{4}$ & $0$ & ${\mathbb Z} \! \times_h \!{\rm D}_4^*$ & $0$ & ${\mathbb Z}$ & ${\mathbb Z}_2$ 
    \\
    \hline
 \end{tabular}
\endgroup
\caption{The nematic phases (a table taken from Ref.~\cite{Masuda:2015jka}).
We show the range of $r$, the phases, the unbroken symmetries $H$, 
the order-parameter manifolds $G/H$, and 
the homotopy groups from $\pi_0$ to $\pi_4$. 
$*$ indicates the universal covering group,  
and ${\mathbb Q} = {\rm D}_2^*$ is a quaternion group. 
For the definition of 
the product $\times_h$, see $\S$4.2.2 and Appendix A of Ref.~\cite{Kobayashi:2011xb}.
}
\label{table-sym}
\end{table*}

We consider the case that the magnetic field is aligned along the $y$ axis, $\vec{B}=(0,B,0)$; 
this can be done without loss of generality.
In fact, it can be shown that this direction gives the most stable state
in the parametrization in Eq.~(\ref{eq:tau_gs}); the other cases in which the magnetic field is aligned along the $x$ or $z$ direction give higher free energy.

\begin{figure}[tb]
\begin{center}
\includegraphics[scale=0.25]{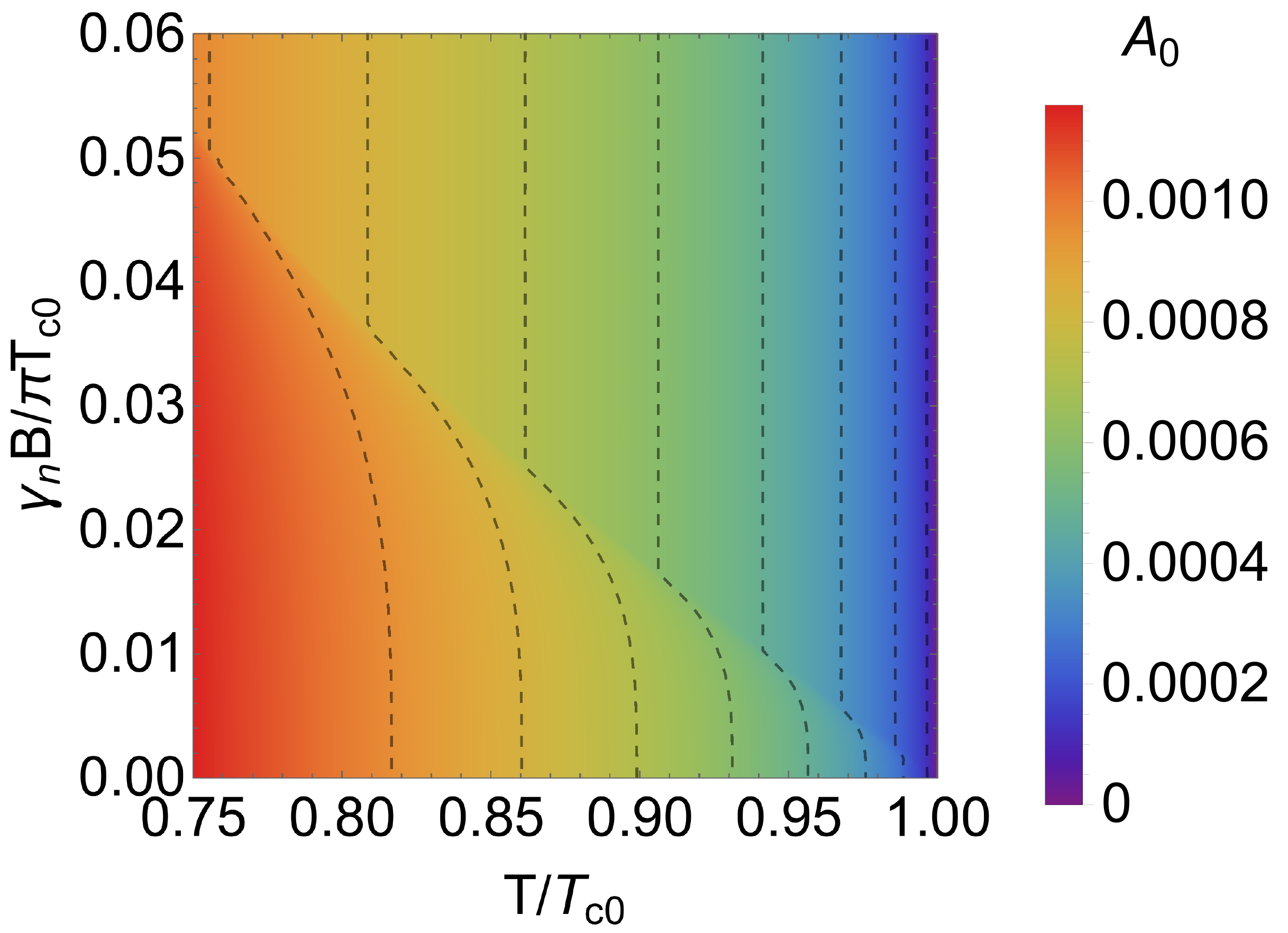}
\hspace{1em}
\includegraphics[scale=0.25]{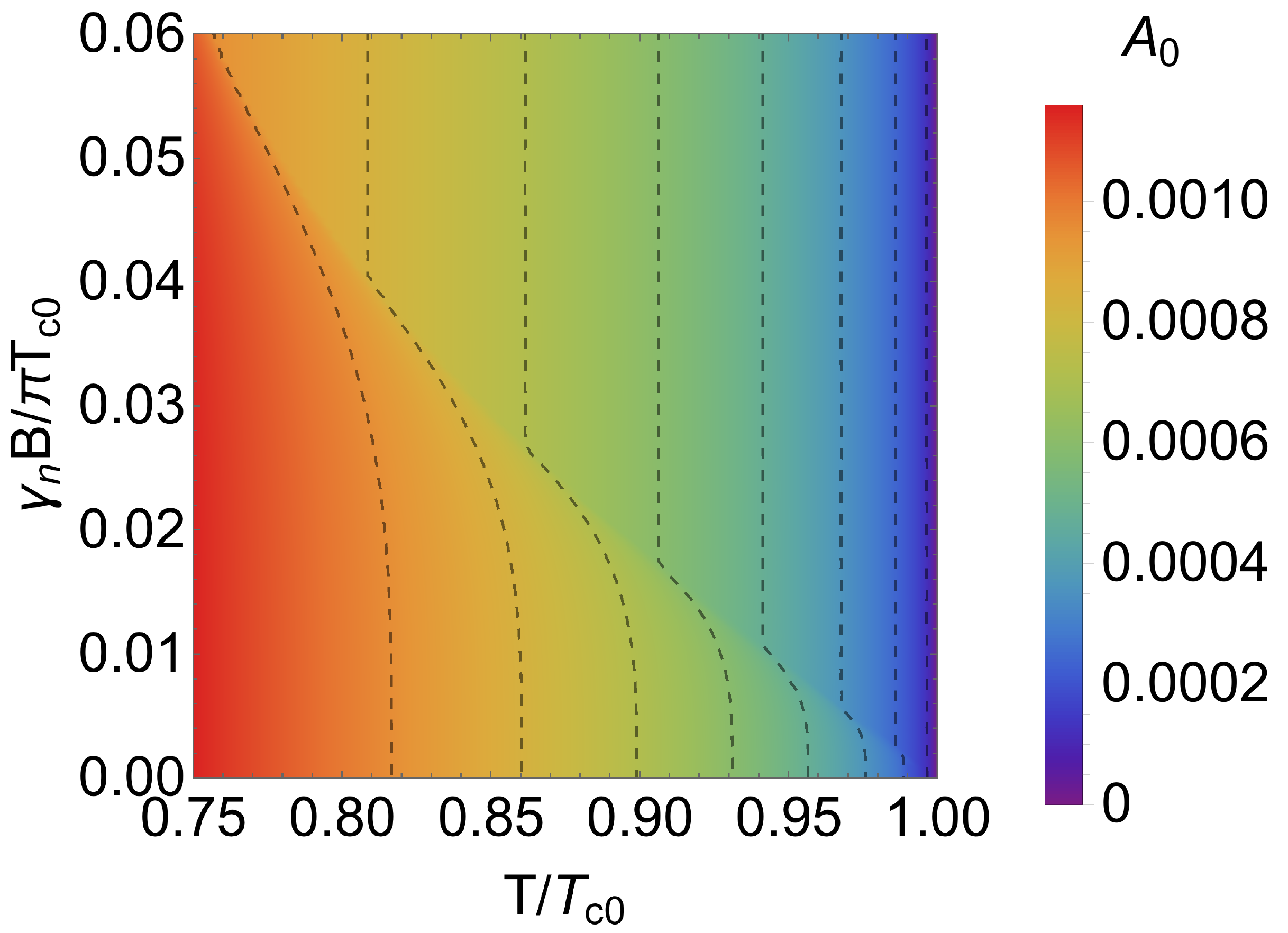}
\\
\vspace{1em}
\includegraphics[scale=0.25]{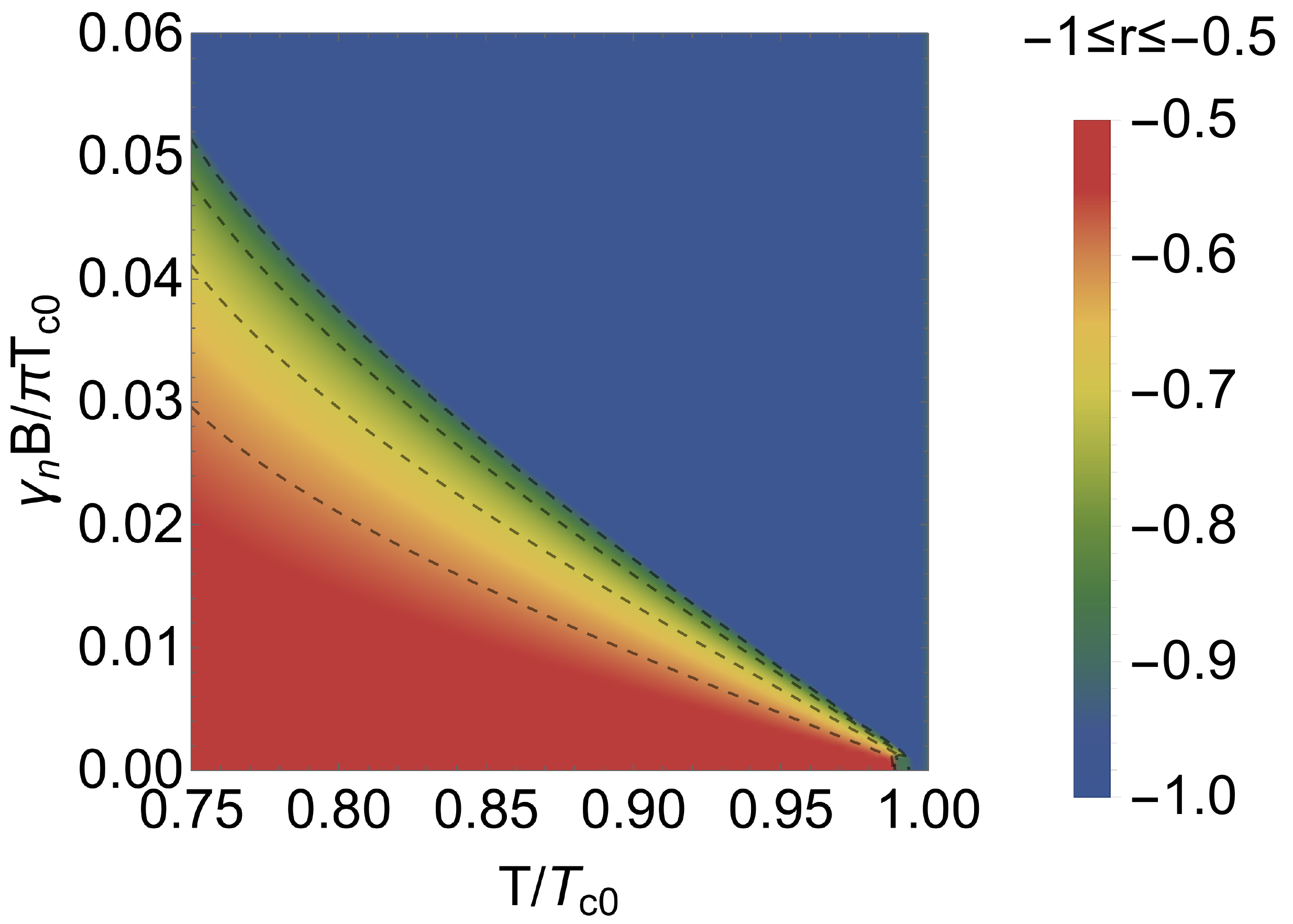}
\hspace{1em}
\includegraphics[scale=0.25]{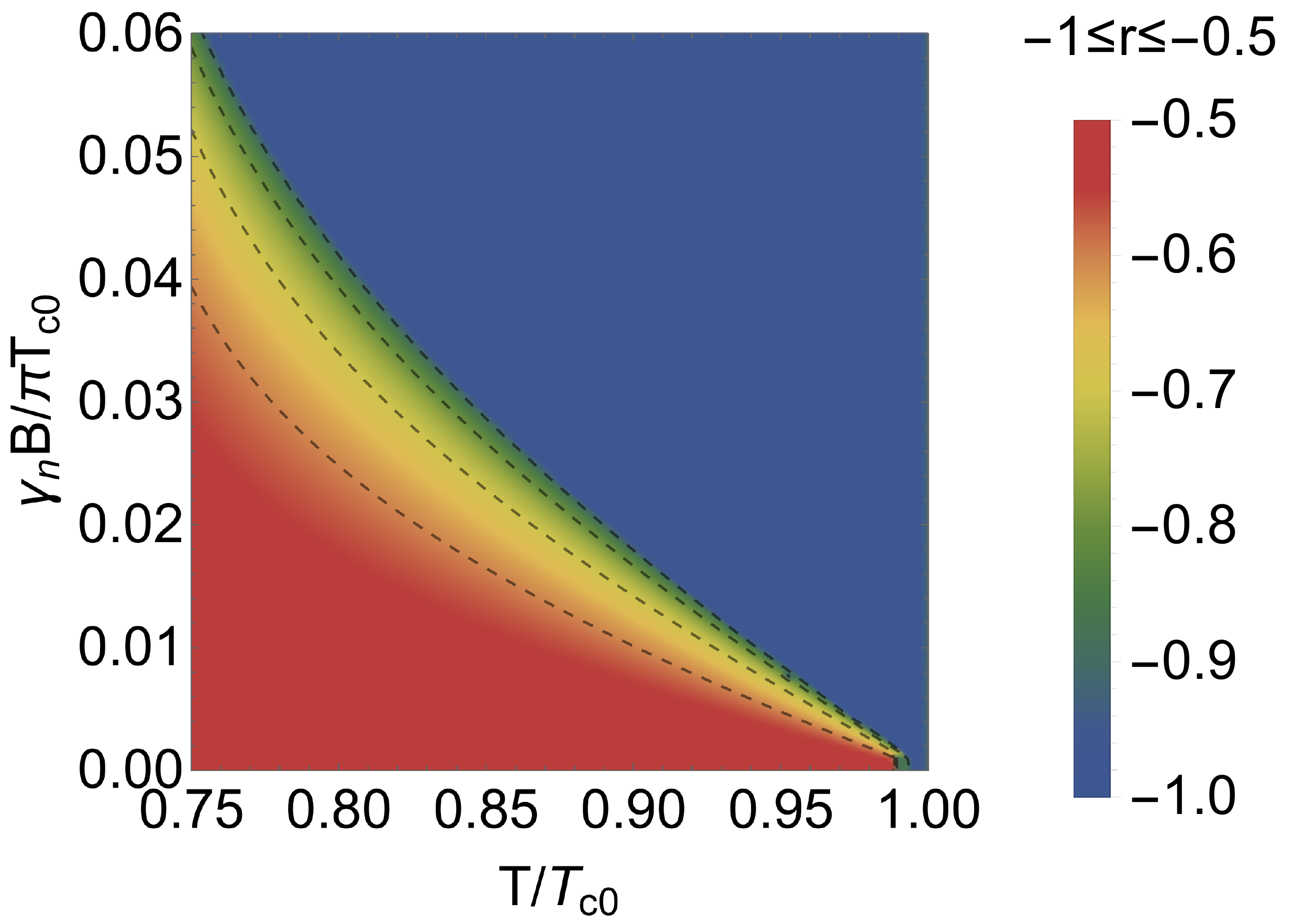}
\caption{The phase diagram for ${A}_{0}$ and $r$ on the $T$-$B$ plane. The two left panels (upper left and  bottom left) are the results up to ${\cal O}(B^{2}{A}^{2})$ in Ref.~\cite{Masuda:2015jka} (setting $\beta^{(4)}=\gamma^{(2)}=0$ in Eq.~(\ref{eq:eff_pot_coefficient02_f})), and the two right panels (upper right and bottom right) are the results up to ${\cal O}(B^{4}{A}^{2})+{\cal O}(B^{2}{A}^{4})$ in the present study. In the bottom panels, the phases are in the UN ($r=-1/2$), D$_{2}$-BN ($-1<r<-1/2$), and D$_{4}$-BN phases ($r=-1$).}
\label{fig:r_B_T_MN_YCN_FL}
\end{center}
\end{figure}

In the above settings, we substitute Eq.~(\ref{eq:tau_gs}) into the GL free energy density in Eq.~(\ref{eq:eff_pot_coefficient02_f}) and perform the variational calculation with respect to ${A}_{0}$ and $r$.
Then, we obtain the phase diagram on the $T$-$B$ plane as shown in Fig.~\ref{fig:r_B_T_MN_YCN_FL}.
We use the dimensionless quantities $T/T_{c0}$ and $\gamma_{n}B/\pi T_{c0}$ for plotting the temperature and the  magnetic field.
Notice that the temperature in the GL free energy should be restricted as $|1-T/T_{c0}| \ll 1$, and also that $B=10^{15}$ [G] ($10^{11}$ [T]), as a typical value in magnetars, corresponds to the value $\gamma_{n}B/\pi T_{c0}=0.02$ for $T_{c0}=0.2$ MeV.
The left panel is the case up to 
${\cal O}({A}^{6})+{\cal O}(B^{2}{A}^{2})$ (i.e., the result by setting $\beta^{(4)}=\gamma^{(2)}=0$ in Eq.~(\ref{eq:eff_pot_coefficient02_f}))~\cite{Masuda:2015jka}, and the right panel is the case up to 
${\cal O}({A}^{6})+{\cal O}(B^{4}{A}^{2})+{\cal O}(B^{2}{A}^{4})$.
In both cases, for $T<T_{c0}$, we observe the following common properties.
At zero magnetic field ($B=0$), 
 the ground state is in the UN phase ($r=-1/2$).\footnote{It is known that the values for $-1\le r \le -1/2$ are degenerate if the terms up to ${\cal O}({A}^{4})$ in $f_{6}^{(0)}$ are included. The degeneracy is resolved when the term of ${\cal O}({A}^{6})$ is included.}
As the magnetic field increases ($B\neq 0$), the phase changes to the D$_{2}$-BN phase ($-1<r<-1/2$) and reaches the D$_{4}$-BN phase ($r=-1$).
All the phase transitions are of the second order.
As a consequence of the effect of ${\cal O}(B^{4}{A}^{2})+{\cal O}(B^{2}{A}^{4})$, the region of the D$_{2}$-BN phase is extended, as shown in the right panel in Fig.~\ref{fig:r_B_T_MN_YCN_FL}.

Notice that the condensate $A_{0}$ in the D$_{4}$-BN phase does not depend on the magnetic field, as shown in the dashed contour lines in Fig.~\ref{fig:r_B_T_MN_YCN_FL}.
This is expected from the energy spectrum of the neutrons in the quasiclassical approximation.
The neutron $^{3}P_{2}$ pairing in the D$_{4}$-BN phase has the spin $\uparrow\uparrow$ component and the spin $\downarrow\downarrow$ component with equal fraction in the D$_{4}$-BN phase, and 
they have energy shifts in the opposite sign due to $-\vec{\mu}_{n}\!\cdot\!\vec{B}$ term in the magnetic field.
Their energy shifts should be canceled in total, because, in the quasiclassical approximation, the higher- and lower-energy states above and below the Fermi surface are treated symmetrically.
This cancellation of the energy shifts turns to be the independence of the condensate $A_{0}$ from the magnetic field in the D$_{4}$-BN phase.

Before concluding our findings about the change in the phase diagram, we need to confirm the validity of the expansion up to 
${\cal O}(B^{4}{A}^{2})+{\cal O}(B^{2}{A}^{4})$ for the magnetic field relevant to neutron stars and magnetars.
To confirm this, we compare the term of ${\cal O}(B^{4}{A}^{2})$ with the term of ${\cal O}(B^{2}{A}^{2})$ in $f_{2}^{(\le4)}$ in Eq.~(\ref{eq:eff_pot_coefficient02_f}).
The ratio is given as
$\displaystyle
   \beta^{(4)}B^{2}/\beta^{(2)} \approx \gamma_{n}^{2}B^{2}/T^{2}
$. 
This quantity should be less than unity in order to achieve a good convergence for the expansion by the magnetic field. 
Considering that the temperature is close to $T_{c0}$ ($T \approx T_{c0}$), we obtain the upper limit of the magnitude of magnetic field: 
 $B \simle T_{c0}/\gamma_{n} = 1.6 \times 10^{16}$ [G] ($1.6 \times 10^{12}$ [T]), i.e., $\gamma_{n}B/(\pi T_{c0}) \simle 0.3$. 
In our analysis based on Eq.~(\ref{eq:eff_pot_coefficient02_f}), in fact, we find that the ground state becomes unstable for stronger magnetic field and no stable solution exists.
Therefore, it is justified that the phase diagrams presented in the right panels in Fig.~\ref{fig:r_B_T_MN_YCN_FL} are within a reasonable range of the magnitude of the magnetic field in the present expansion.

\subsection{Thermodynamic properties}

\begin{figure}[tb]
\begin{center}
\includegraphics[scale=0.25]{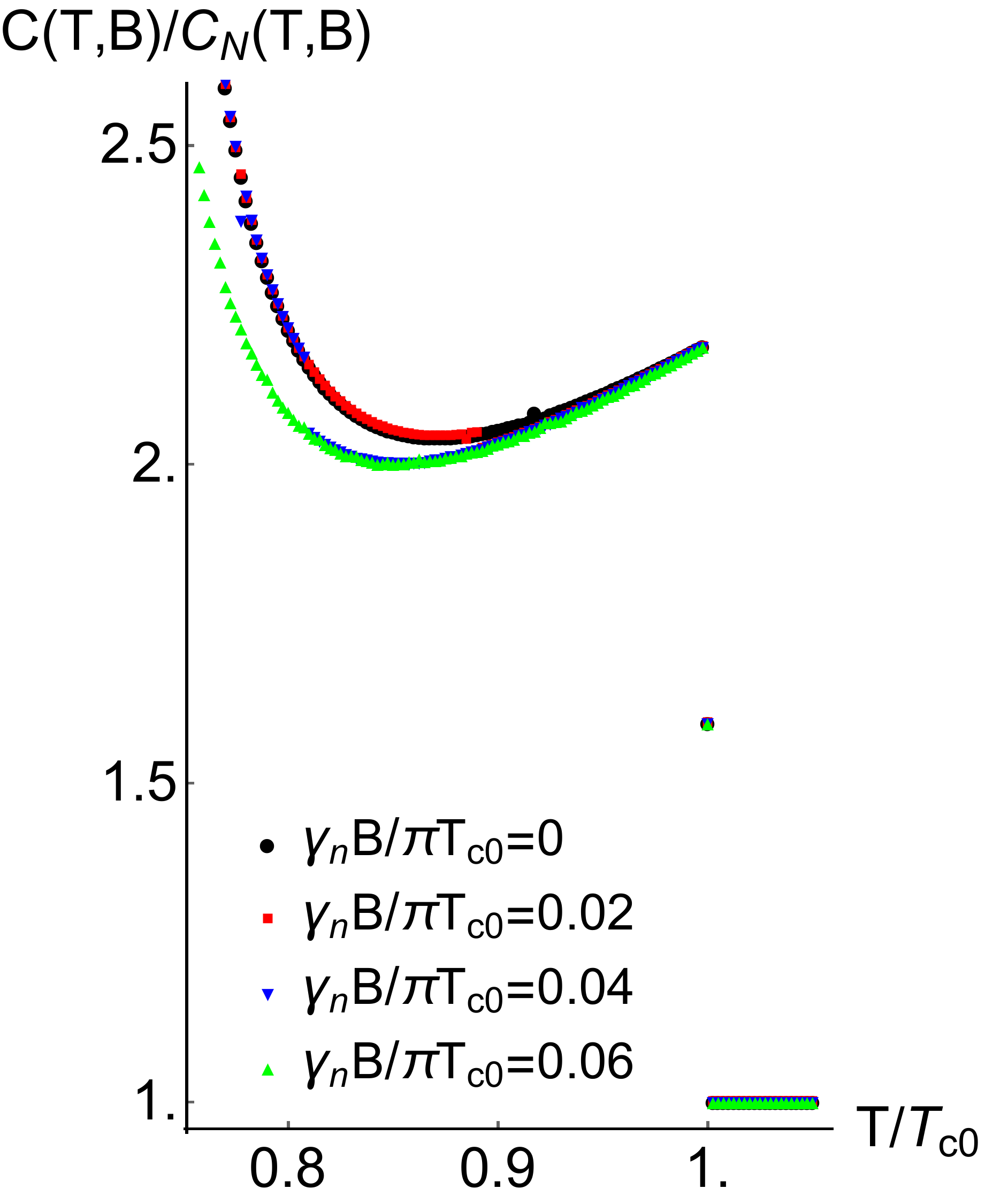}
\hspace{1em}
\includegraphics[scale=0.25]{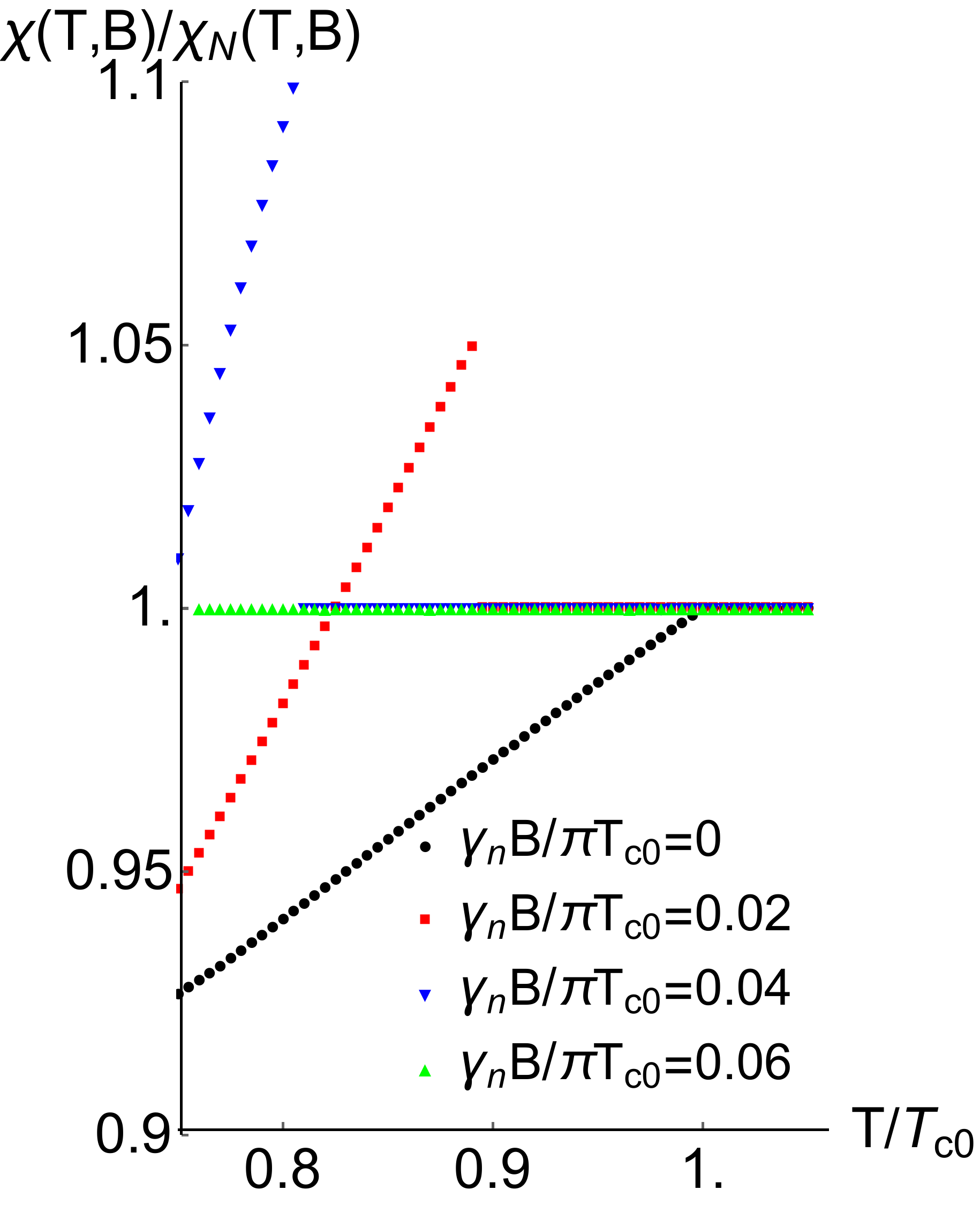}
\caption{The heat capacity $C(T,B)$ and the magnetic susceptibility $\chi(T,B)=\chi_{2}(T,B)$ along the direction parallel to the magnetic field are plotted. They are normalized by $C_{N}(T,B)$ and $\chi_{N}(T,B)$, respectively.}
\label{fig:C_chi_T_YCN_FL}
\end{center}
\end{figure}

We investigate the thermodynamic properties, heat capacity and magnetic susceptibility, of the neutron $^{3}P_{2}$ superfluid.
They will be useful quantities for the investigation of the inner structures in neutron stars by observations.
The heat capacity is defined by
\begin{eqnarray}
   C(T,B) = -T \frac{\partial^{2}f}{\partial T^{2}} \biggr|_{\vec{B}=(0,B,0)}.
\end{eqnarray}
We plot the heat capacity normalized by the heat capacity of a free neutron gas, $C_{N}(T,B) =  (2\pi^{2}/3) N(0) T$, in the left panel in Fig.~\ref{fig:C_chi_T_YCN_FL}.
The spin susceptibility along the spatial direction $i=1,2,3$ is defined by
\begin{eqnarray}
   \chi_{i}(T,B) = \frac{\partial M_{i}(T,B)}{\partial B_{i}} \biggr|_{\vec{B}=(0,B,0)}, \hspace{1em}
\label{eq:spin_susceptibility_definition_2}
\end{eqnarray}
with the magnetization 
$\displaystyle M_{i}(T,B) = -(1+F_{0}^{a}){\partial f}/{\partial B_{i}} \bigr|_{\vec{B}=(0,B,0)}$.
Notice that the direction of the magnetic field is fixed to be along the $y$ axis, $\vec{B}=(0,B,0)$.
We plot $\chi(T,B)=\chi_{2}(T,B)$, which is directed along the $y$ axis and is normalized by the spin susceptibility for a free neutron gas, $\chi_{N}(T,B) = 2N(0)\mu_{\!n}^{2}/(1+F_{0}^{a})$, in the right panel in Fig.~\ref{fig:C_chi_T_YCN_FL}.
Both in the heat capacity and in the spin susceptibility, we find some discontinuities, which indicate that the phase transition is of second order.
In the left panel of the figure, the heat capacity jumps drastically at the temperature $T \approx T_{c0}$.
This is the phase transition from the $^{3}P_{2}$ superfluid phase to the normal phase.
At finite magnetic field, we find small jumps at $T=0.89T_{c0}$ for $\gamma_{n}B/\pi T_{c0}=0.02$ and $T=0.81T_{c0}$ for $\gamma_{n}B/\pi T_{c0}=0.04$.
Similarly, we also find jumps in the spin susceptibility, as shown in the right panel of the figure.
They are at phase transitions of the second order from the D$_{2}$-BN and D$_{4}$-BN phases (cf.~Fig.~\ref{fig:r_B_T_MN_YCN_FL}).
We may notice that the heat capacity has a minimum as a function of temperature in the left panel.
However, this should not be considered seriously because the temperature region should be restricted to $|1-T/T_{c0}| \ll 1$.
In the present parameter set, the temperature region with $T/T_{c0} \gtrsim 0.8$ would be reasonably acceptable.
Our result is consistent with the result in the analysis by the BdG equation that the heat capacity decreases as the temperature decreases in the superfluid phase~\cite{Mizushima:2016fbn}.
As for the spin susceptibility, we notice that $\chi_{2}$ in the D$_{4}$-BN phase has no change from that of a free neutron gas.
This is expected from the energy spectrum of the neutrons in the magnetic field, as discussed previously.
Thus, the response of the neutron $^{3}P_{2}$ superfluidity to the magnetic field seems to be the same as that of a free neutron gas.
However, the spin susceptibilities, $\chi_{1}$ and $\chi_{3}$, in the direction perpendicular to the applied magnetic field are still dependent on the magnetic field.
Thus, there exists the nontrivial magnetic response
 in the D$_{4}$-BN phase.

\begin{figure}[tb]
\begin{center}
\includegraphics[scale=0.25]{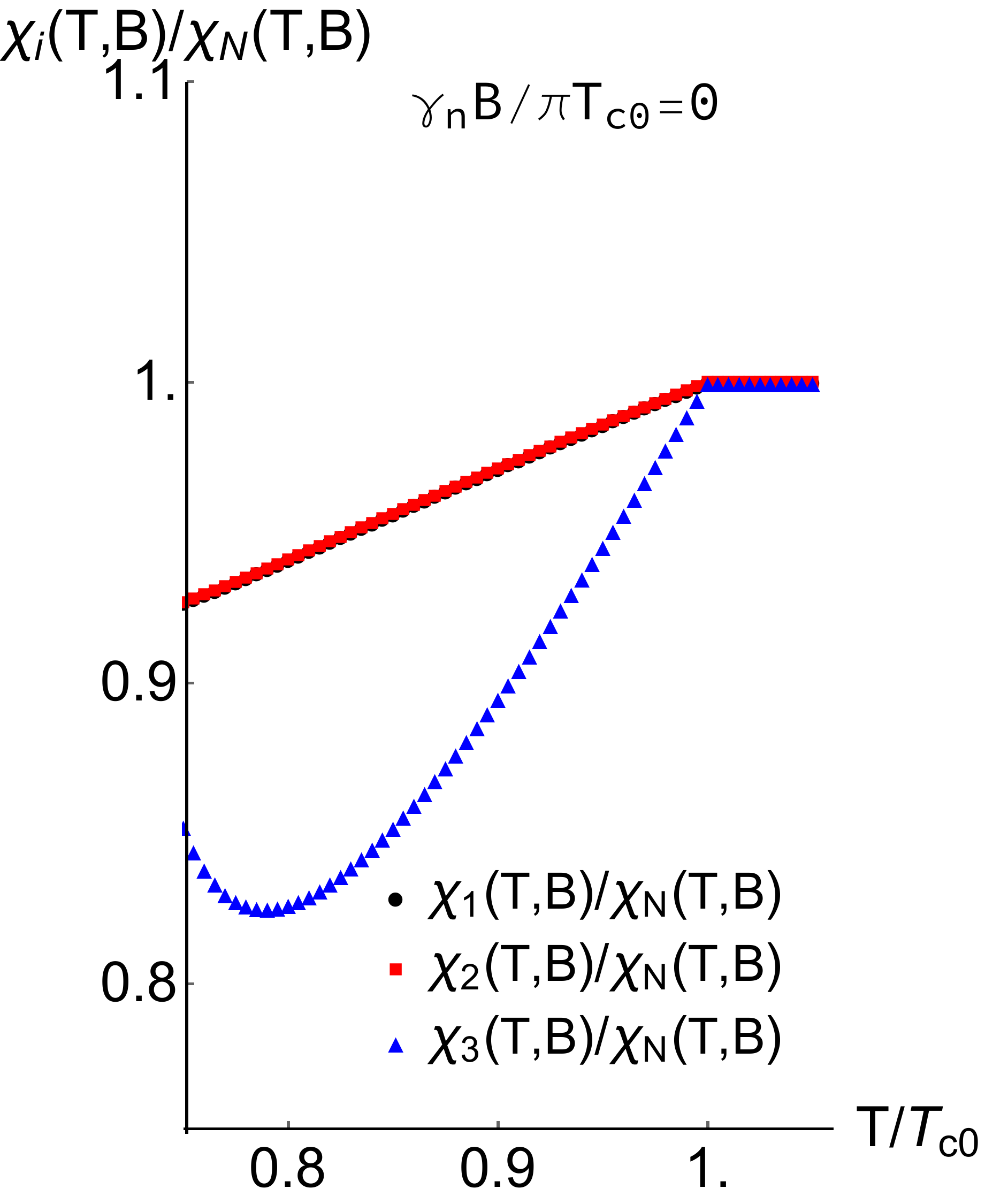}
\hspace{1em}
\includegraphics[scale=0.25]{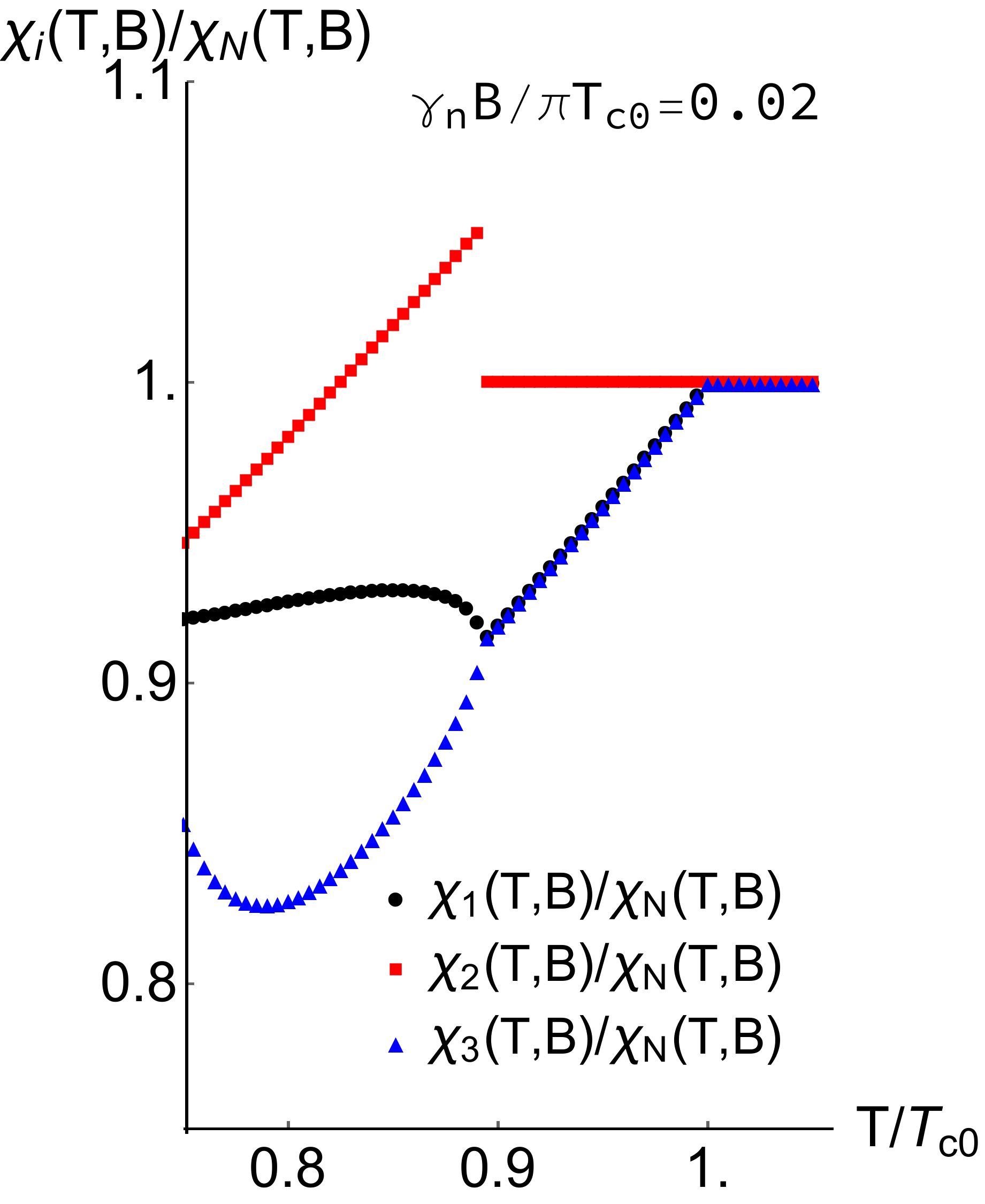}
\\
\vspace{2em}
\includegraphics[scale=0.25]{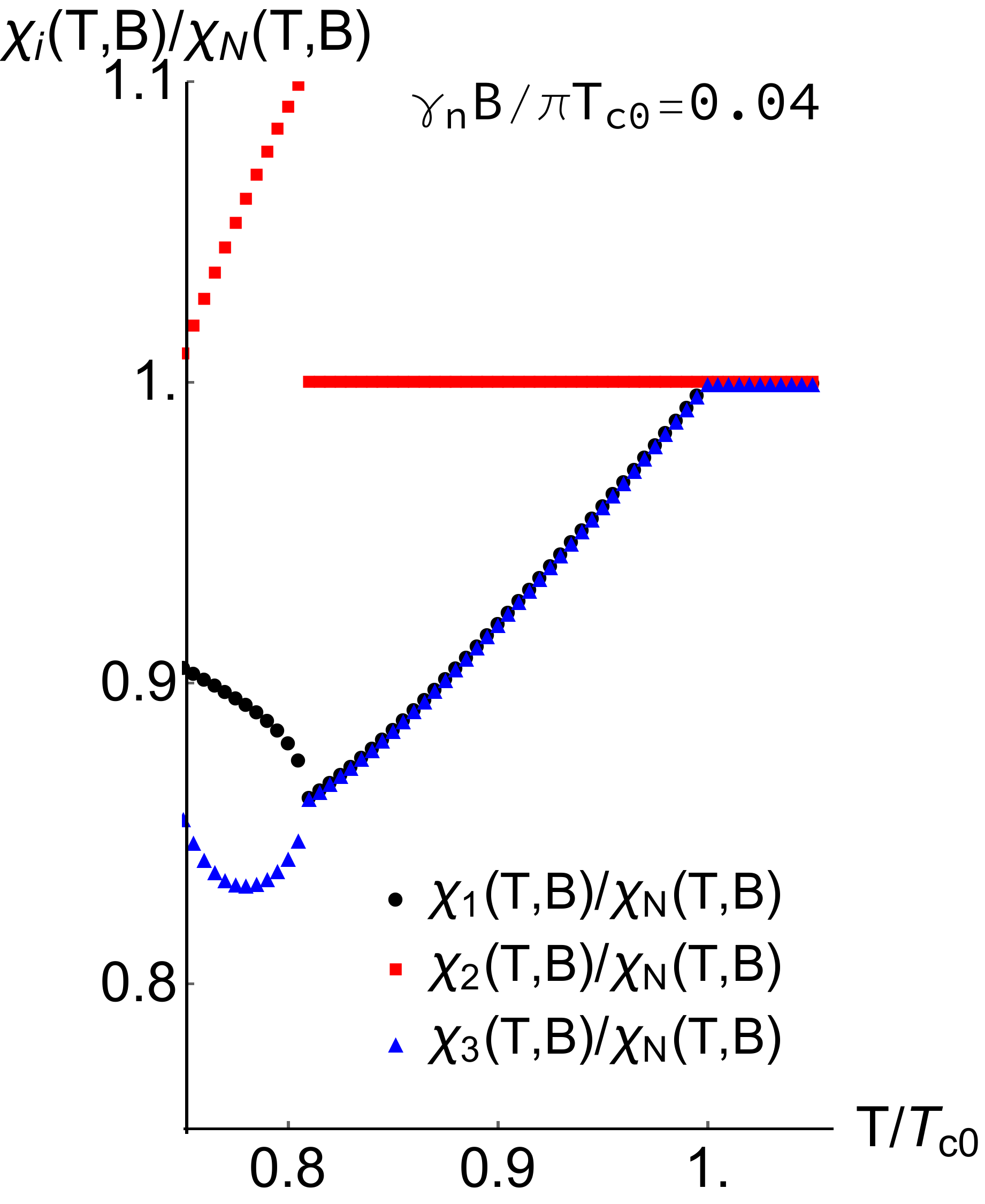}
\hspace{1em}
\includegraphics[scale=0.25]{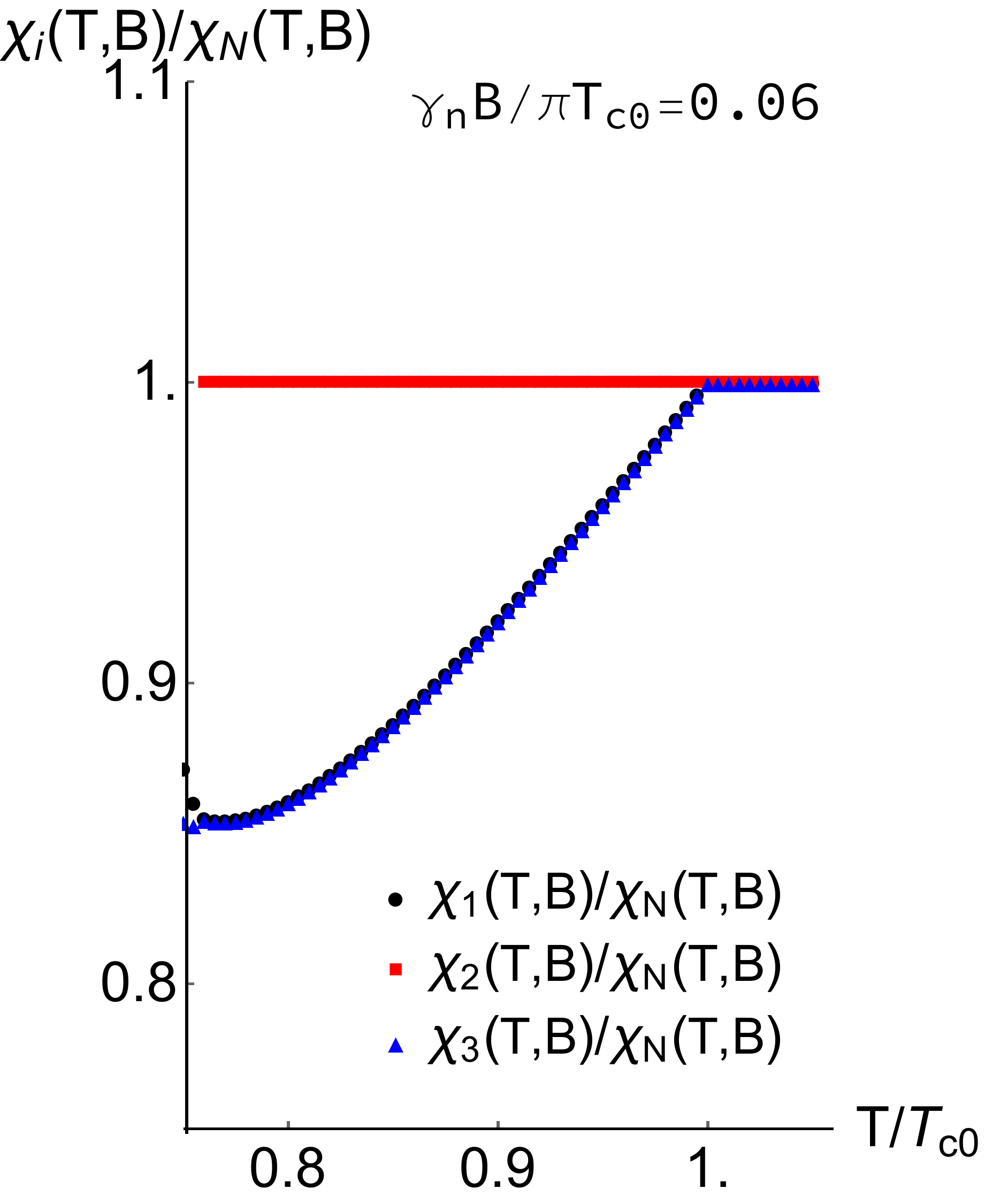}
\caption{Spin susceptibilities  $\chi_{i}(T,B)$ ($i=1,2,3$) for various magnitudes of magnetic field. They are normalized by $\chi_{N}(T,B)$.}
\label{fig:chi123_T_YCN_FL_B}
\end{center}
\end{figure}

Finally, we discuss the (an)isotropy of the spin susceptibility. 
We plot $\chi_{i}(B,T)$ ($i=1,2,3$) for the fixed magnetic field $\vec{B}=(0,B,0)$ in Fig.~\ref{fig:chi123_T_YCN_FL_B}.
We find that $\chi_{1}$, $\chi_{2}$ and, $\chi_{3}$ exhibit different behavior in the superfluid phase.
It is interesting to note that $\chi_{1}$ and $\chi_{2}$ are the same (isotropic) in the UN phase and are different (anisotropic) in the D$_{2}$-BN and D$_{4}$-BN phases. 
Similarly, $\chi_{1}$ and $\chi_{3}$ are different (anisotropic) in the UN and D$_{2}$-BN phases and are the same (isotropic) in the D$_{4}$-BN phase (cf.~Fig.~\ref{fig:r_B_T_MN_YCN_FL}).
Such (an)isotropy can be understood by the energy-momentum dispersion relations for neutrons in the UN, D$_{2}$-BN, and D$_{4}$-BN phases.
From the neutron propagator (\ref{eq:Sm_propagator_def}) and the vertex function (\ref{eq:hat_tau_def}), we obtain the energy-momentum dispersion relation of the neutron in the $^{3}P_{2}$ phase,
\begin{eqnarray}
  E_{\pm}({A}_{0},r)
=
\pm
\sqrt{
          \frac{\bar{\gamma}_{n}^{2}}{4} \vec{B}^{2} + \vec{d}\!\cdot\!\vec{d}^{\ast} + \xi_{\vec{p}}^{2}
   \pm \sqrt{
              - 2 |\vec{d}\times\vec{d}^{\ast}|^{2}
             + \bar{\gamma}_{n}^{2} (\vec{B}\!\cdot\!\vec{d}) (\vec{B}\!\cdot\!\vec{d}^{\ast})
             - 2i \bar{\gamma}_{n} \vec{B}\!\cdot\!(\vec{d}\times\vec{d}^{\ast}) \xi_{\vec{p}}
             + \bar{\gamma}_{n}^{2} \vec{B}^{2} \xi_{\vec{p}}^{2}
          } 
},
\end{eqnarray}
with $\vec{d}={A}\vec{p}$ and three-dimensional momentum $\vec{p}=(p_{1},p_{2},p_{3})$ of a neutron.
In the case of $\vec{B}=(0,B,0)$, the above equation becomes
\begin{eqnarray}
   E_{\pm}({A}_{0},r) \bigl|_{\vec{B}=(0,B,0)}
= \pm
\sqrt{
\frac{\bar{\gamma}_{n}^{2}}{4}B^{2}
+{A}_{0}^{2} \bigl( r^{2} p_{1}^{2} + (1+r)^{2} p_{2}^{2} + p_{3}^{2} \bigr)+\xi_{\vec{p}}^{2}
\pm \bar{\gamma}_{n} B \sqrt{{A}_{0}^{2} (1+r)^{2} p_{2}^{2}+\xi_{\vec{p}}^{2}}
}.
\end{eqnarray}
We observe that the energy-momentum dispersion is anisotropic on the $(p_{1},p_{3})$ plane, perpendicular to the $y$ axis, for $-1<r<-1/2$ (the D$_{2}$-BN phase), and that it is isotropic on the $(p_{1},p_{2})$ plane for $r=-1/2$ (the UN phase) and it is isotropic also on the $(p_{1},p_{3})$-plane for $r=-1$ (the D$_{4}$-BN phase).
Thus, we confirm that the (an)isotropy of the energy-momentum dispersion relation for the neutron in the $^{3}P_{2}$ phase is correctly reflected in the GL free-energy density.

\section{Summary and discussion}
\label{sec:conclusion}

We have studied the effect of a strong magnetic field on the neutron $^{3}P_{2}$ superfluidity.
Assuming the attraction in the neutron $^{3}P_{2}$ pair, as expected from the $LS$ interaction,
we have derived the GL free energy by integrating out the neutron fields at the one-loop level and by adopting the low-energy approximation near 
the critical temperature.
We have kept the next-to-leading-order terms of the magnetic field in the expansion of the GL free energy.
It has been found that the region of the D$_{2}$-BN phase is extended at strong magnetic fields, as the effect of next-to-leading order in the expansion for the magnetic field.
We also investigate the thermodynamic properties, such as the heat capacity and spin susceptibility, and show that the spin susceptibility 
 is anisotropic in the D$_{2}$-BN phase and isotropic in the UN and D$_{4}$-BN phases, depending on the direction relative to the magnetic field.
 Such the (an)isotropy of the spin susceptibility may help us to understand the neutron $^{3}P_{2}$ superfluids inside magnetars in observations.

The phase transitions are of the second order 
 in the GL analysis in this paper. 
The phase diagram derived from the BdG equation 
contains a first-order transition between the D$_2$-BN and D$_4$-BN phases for lower temperature and the tricritical point between the first and second-order transition lines~\cite{Mizushima:2016fbn}.
These may be incorporated in the GL analysis 
by taking into account higher-order term in ${A}$.
Beyond the particle-hole symmetry, it is an interesting question to ask what is the change of the phase diagram in the particle-hole asymmetry, i.e., the competition between the nematic phase and the ferromagnetic phase.

For future work, 
it is an interesting question to ask how the neutron $^{3}P_{2}$ phase is connected to the other phases, such as hyperon matter, quark matter, and so on.
For example, the connection of the (non-Abelian) quantum vortices between the hadron matter and the quark matter is discussed in the view of boojums~\cite{Cipriani:2012hr,Alford:2018mqj,Chatterjee:2018nxe,Cherman:2018jir}.\footnote{See Ref.~\cite{Eto:2013hoa} for a review of non-Abelian quantum vortices in quark matter.}
The applications to the observables in the magnetars are also important.

It is also interesting to investigate a possibility of realization of different phases of  $^{3}P_{2}$ superfluids. 
In the Mermin's classification in the GL theory, there can exist 
cyclic and ferromagnetic phases. 
The former admits 1/3 quantized non-Abelian vortices~\cite{Semenoff:2006vv}
forming a network in collision~\cite{Kobayashi:2008pk}, 
and the latter could be the origin of strong magnetic field of magnetars.
Furthermore, 
both phases have been found to be topological matter, 
called Weyl semimetals, 
admitting gapless fermions in the bulk~\cite{Mizushima:2016fbn,Mizushima:2017pma},
which would significantly affect the cooling properties of neutron stars.
Applications to higher spin systems will be also interesting~\cite{PhysRevX.8.011029}.
Those subjects are left for future works.

\section*{Acknowledgment}
We would like to thank Motoi Tachibana for discussions.
We would like to thank Yusuke Masaki for important comments.
This work is supported by the Ministry of Education, Culture, Sports, Science (MEXT)-Supported Program for the Strategic Research Foundation at Private Universities ``Topological Science" (Grant No. S1511006). 
C.~C. acknowledges support as an International Research Fellow of the Japan Society for the Promotion of Science (JSPS) (Grant No. 16F16322). 
This work is also supported in part by 
JSPS Grant-in-Aid for Scientific Research (KAKENHI Grant No. 16H03984 (M.~N.), No.~18H01217 (M.~N.) and, No.~17K05435 (S.~Y.)), and also by MEXT KAKENHI Grant-in-Aid for Scientific Research on Innovative Areas ``Topological Materials Science'' No.~15H05855 (M.~N.).

\bibliography{reference}

\begin{thebibliography}{82}%
\makeatletter
\providecommand \@ifxundefined [1]{%
 \@ifx{#1\undefined}
}%
\providecommand \@ifnum [1]{%
 \ifnum #1\expandafter \@firstoftwo
 \else \expandafter \@secondoftwo
 \fi
}%
\providecommand \@ifx [1]{%
 \ifx #1\expandafter \@firstoftwo
 \else \expandafter \@secondoftwo
 \fi
}%
\providecommand \natexlab [1]{#1}%
\providecommand \enquote  [1]{``#1''}%
\providecommand \bibnamefont  [1]{#1}%
\providecommand \bibfnamefont [1]{#1}%
\providecommand \citenamefont [1]{#1}%
\providecommand \href@noop [0]{\@secondoftwo}%
\providecommand \href [0]{\begingroup \@sanitize@url \@href}%
\providecommand \@href[1]{\@@startlink{#1}\@@href}%
\providecommand \@@href[1]{\endgroup#1\@@endlink}%
\providecommand \@sanitize@url [0]{\catcode `\\12\catcode `\$12\catcode
  `\&12\catcode `\#12\catcode `\^12\catcode `\_12\catcode `\%12\relax}%
\providecommand \@@startlink[1]{}%
\providecommand \@@endlink[0]{}%
\providecommand \url  [0]{\begingroup\@sanitize@url \@url }%
\providecommand \@url [1]{\endgroup\@href {#1}{\urlprefix }}%
\providecommand \urlprefix  [0]{URL }%
\providecommand \Eprint [0]{\href }%
\providecommand \doibase [0]{http://dx.doi.org/}%
\providecommand \selectlanguage [0]{\@gobble}%
\providecommand \bibinfo  [0]{\@secondoftwo}%
\providecommand \bibfield  [0]{\@secondoftwo}%
\providecommand \translation [1]{[#1]}%
\providecommand \BibitemOpen [0]{}%
\providecommand \bibitemStop [0]{}%
\providecommand \bibitemNoStop [0]{.\EOS\space}%
\providecommand \EOS [0]{\spacefactor3000\relax}%
\providecommand \BibitemShut  [1]{\csname bibitem#1\endcsname}%
\let\auto@bib@innerbib\@empty
\bibitem [{\citenamefont {Graber}\ \emph {et~al.}(2017)\citenamefont {Graber},
  \citenamefont {Andersson},\ and\ \citenamefont {Hogg}}]{Graber:2016imq}%
  \BibitemOpen
  \bibfield  {author} {\bibinfo {author} {\bibfnamefont {V.}~\bibnamefont
  {Graber}}, \bibinfo {author} {\bibfnamefont {N.}~\bibnamefont {Andersson}}, \
  and\ \bibinfo {author} {\bibfnamefont {M.}~\bibnamefont {Hogg}},\ }\href
  {\doibase 10.1142/S0218271817300154} {\bibfield  {journal} {\bibinfo
  {journal} {Int. J. Mod. Phys.}\ }\textbf {\bibinfo {volume} {D26}},\ \bibinfo
  {pages} {1730015} (\bibinfo {year} {2017})},\ \Eprint
  {http://arxiv.org/abs/1610.06882} {arXiv:1610.06882 [astro-ph.HE]}
  \BibitemShut {NoStop}%
\bibitem [{\citenamefont {Baym}\ \emph {et~al.}(2018)\citenamefont {Baym},
  \citenamefont {Hatsuda}, \citenamefont {Kojo}, \citenamefont {Powell},
  \citenamefont {Song},\ and\ \citenamefont {Takatsuka}}]{Baym:2017whm}%
  \BibitemOpen
  \bibfield  {author} {\bibinfo {author} {\bibfnamefont {G.}~\bibnamefont
  {Baym}}, \bibinfo {author} {\bibfnamefont {T.}~\bibnamefont {Hatsuda}},
  \bibinfo {author} {\bibfnamefont {T.}~\bibnamefont {Kojo}}, \bibinfo {author}
  {\bibfnamefont {P.~D.}\ \bibnamefont {Powell}}, \bibinfo {author}
  {\bibfnamefont {Y.}~\bibnamefont {Song}}, \ and\ \bibinfo {author}
  {\bibfnamefont {T.}~\bibnamefont {Takatsuka}},\ }\href {\doibase
  10.1088/1361-6633/aaae14} {\bibfield  {journal} {\bibinfo  {journal} {Rept.
  Prog. Phys.}\ }\textbf {\bibinfo {volume} {81}},\ \bibinfo {pages} {056902}
  (\bibinfo {year} {2018})},\ \Eprint {http://arxiv.org/abs/1707.04966}
  {arXiv:1707.04966 [astro-ph.HE]} \BibitemShut {NoStop}%
\bibitem [{\citenamefont {Abbott}\ \emph {et~al.}(2017)\citenamefont {Abbott}
  \emph {et~al.}}]{TheLIGOScientific:2017qsa}%
  \BibitemOpen
  \bibfield  {author} {\bibinfo {author} {\bibfnamefont {B.}~\bibnamefont
  {Abbott}} \emph {et~al.} (\bibinfo {collaboration} {Virgo, LIGO
  Scientific}),\ }\href {\doibase 10.1103/PhysRevLett.119.161101} {\bibfield
  {journal} {\bibinfo  {journal} {Phys. Rev. Lett.}\ }\textbf {\bibinfo
  {volume} {119}},\ \bibinfo {pages} {161101} (\bibinfo {year} {2017})},\
  \Eprint {http://arxiv.org/abs/1710.05832} {arXiv:1710.05832 [gr-qc]}
  \BibitemShut {NoStop}%
\bibitem [{\citenamefont {Turolla}\ \emph {et~al.}(2015)\citenamefont
  {Turolla}, \citenamefont {Zane},\ and\ \citenamefont
  {Watts}}]{Turolla:2015mwa}%
  \BibitemOpen
  \bibfield  {author} {\bibinfo {author} {\bibfnamefont {R.}~\bibnamefont
  {Turolla}}, \bibinfo {author} {\bibfnamefont {S.}~\bibnamefont {Zane}}, \
  and\ \bibinfo {author} {\bibfnamefont {A.}~\bibnamefont {Watts}},\ }\href
  {\doibase 10.1088/0034-4885/78/11/116901} {\bibfield  {journal} {\bibinfo
  {journal} {Rept. Prog. Phys.}\ }\textbf {\bibinfo {volume} {78}},\ \bibinfo
  {pages} {116901} (\bibinfo {year} {2015})},\ \Eprint
  {http://arxiv.org/abs/1507.02924} {arXiv:1507.02924 [astro-ph.HE]}
  \BibitemShut {NoStop}%
\bibitem [{\citenamefont {Kaspi}\ and\ \citenamefont
  {Beloborodov}(2017)}]{Kaspi2017}%
  \BibitemOpen
  \bibfield  {author} {\bibinfo {author} {\bibfnamefont {V.~M.}\ \bibnamefont
  {Kaspi}}\ and\ \bibinfo {author} {\bibfnamefont {A.~M.}\ \bibnamefont
  {Beloborodov}},\ }\href {\doibase 10.1146/annurev-astro-081915-023329}
  {\bibfield  {journal} {\bibinfo  {journal} {Annual Review of Astronomy and
  Astrophysics}\ }\textbf {\bibinfo {volume} {55}},\ \bibinfo {pages} {261}
  (\bibinfo {year} {2017})}\BibitemShut {NoStop}%
\bibitem [{\citenamefont {Brownell}\ and\ \citenamefont
  {Callaway}(1969)}]{Brownell1969}%
  \BibitemOpen
  \bibfield  {author} {\bibinfo {author} {\bibfnamefont {D.~H.}\ \bibnamefont
  {Brownell}}\ and\ \bibinfo {author} {\bibfnamefont {J.}~\bibnamefont
  {Callaway}},\ }\href {\doibase 10.1007/BF02712360} {\bibfield  {journal}
  {\bibinfo  {journal} {Il Nuovo Cimento B (1965-1970)}\ }\textbf {\bibinfo
  {volume} {60}},\ \bibinfo {pages} {169} (\bibinfo {year} {1969})}\BibitemShut
  {NoStop}%
\bibitem [{\citenamefont {Rice}(1969)}]{RICE1969637}%
  \BibitemOpen
  \bibfield  {author} {\bibinfo {author} {\bibfnamefont {M.}~\bibnamefont
  {Rice}},\ }\href {\doibase https://doi.org/10.1016/0375-9601(69)91141-4}
  {\bibfield  {journal} {\bibinfo  {journal} {Physics Letters A}\ }\textbf
  {\bibinfo {volume} {29}},\ \bibinfo {pages} {637 } (\bibinfo {year}
  {1969})}\BibitemShut {NoStop}%
\bibitem [{\citenamefont {Silverstein}(1969)}]{Silverstein:1969zz}%
  \BibitemOpen
  \bibfield  {author} {\bibinfo {author} {\bibfnamefont {S.~D.}\ \bibnamefont
  {Silverstein}},\ }\href {\doibase 10.1103/PhysRevLett.23.139} {\bibfield
  {journal} {\bibinfo  {journal} {Phys. Rev. Lett.}\ }\textbf {\bibinfo
  {volume} {23}},\ \bibinfo {pages} {139} (\bibinfo {year} {1969})}\BibitemShut
  {NoStop}%
\bibitem [{\citenamefont {Haensel}\ and\ \citenamefont
  {Bonazzola}(1996)}]{Haensel:1996ss}%
  \BibitemOpen
  \bibfield  {author} {\bibinfo {author} {\bibfnamefont {P.}~\bibnamefont
  {Haensel}}\ and\ \bibinfo {author} {\bibfnamefont {S.}~\bibnamefont
  {Bonazzola}},\ }\href@noop {} {\bibfield  {journal} {\bibinfo  {journal}
  {Astron. Astrophys.}\ }\textbf {\bibinfo {volume} {314}},\ \bibinfo {pages}
  {1017} (\bibinfo {year} {1996})},\ \Eprint
  {http://arxiv.org/abs/astro-ph/9605149} {arXiv:astro-ph/9605149 [astro-ph]}
  \BibitemShut {NoStop}%
\bibitem [{\citenamefont {Bordbar}\ and\ \citenamefont
  {Bigdeli}(2008)}]{Bordbar:2008zz}%
  \BibitemOpen
  \bibfield  {author} {\bibinfo {author} {\bibfnamefont {G.~H.}\ \bibnamefont
  {Bordbar}}\ and\ \bibinfo {author} {\bibfnamefont {M.}~\bibnamefont
  {Bigdeli}},\ }\href {\doibase 10.1103/PhysRevC.77.015805} {\bibfield
  {journal} {\bibinfo  {journal} {Phys. Rev.}\ }\textbf {\bibinfo {volume}
  {C77}},\ \bibinfo {pages} {015805} (\bibinfo {year} {2008})},\ \Eprint
  {http://arxiv.org/abs/0809.3498} {arXiv:0809.3498 [nucl-th]} \BibitemShut
  {NoStop}%
\bibitem [{\citenamefont {Eto}\ \emph {et~al.}(2013)\citenamefont {Eto},
  \citenamefont {Hashimoto},\ and\ \citenamefont {Hatsuda}}]{Eto:2012qd}%
  \BibitemOpen
  \bibfield  {author} {\bibinfo {author} {\bibfnamefont {M.}~\bibnamefont
  {Eto}}, \bibinfo {author} {\bibfnamefont {K.}~\bibnamefont {Hashimoto}}, \
  and\ \bibinfo {author} {\bibfnamefont {T.}~\bibnamefont {Hatsuda}},\ }\href
  {\doibase 10.1103/PhysRevD.88.081701} {\bibfield  {journal} {\bibinfo
  {journal} {Phys. Rev.}\ }\textbf {\bibinfo {volume} {D88}},\ \bibinfo {pages}
  {081701} (\bibinfo {year} {2013})},\ \Eprint {http://arxiv.org/abs/1209.4814}
  {arXiv:1209.4814 [hep-ph]} \BibitemShut {NoStop}%
\bibitem [{\citenamefont {Hashimoto}(2015)}]{Hashimoto:2014sha}%
  \BibitemOpen
  \bibfield  {author} {\bibinfo {author} {\bibfnamefont {K.}~\bibnamefont
  {Hashimoto}},\ }\href {\doibase 10.1103/PhysRevD.91.085013} {\bibfield
  {journal} {\bibinfo  {journal} {Phys. Rev.}\ }\textbf {\bibinfo {volume}
  {D91}},\ \bibinfo {pages} {085013} (\bibinfo {year} {2015})},\ \Eprint
  {http://arxiv.org/abs/1412.6960} {arXiv:1412.6960 [hep-ph]} \BibitemShut
  {NoStop}%
\bibitem [{\citenamefont {Tatsumi}(2000)}]{Tatsumi:1999ab}%
  \BibitemOpen
  \bibfield  {author} {\bibinfo {author} {\bibfnamefont {T.}~\bibnamefont
  {Tatsumi}},\ }\href {\doibase 10.1016/S0370-2693(00)00927-8} {\bibfield
  {journal} {\bibinfo  {journal} {Phys. Lett.}\ }\textbf {\bibinfo {volume}
  {B489}},\ \bibinfo {pages} {280} (\bibinfo {year} {2000})},\ \Eprint
  {http://arxiv.org/abs/hep-ph/9910470} {arXiv:hep-ph/9910470 [hep-ph]}
  \BibitemShut {NoStop}%
\bibitem [{\citenamefont {Nakano}\ \emph {et~al.}(2003)\citenamefont {Nakano},
  \citenamefont {Maruyama},\ and\ \citenamefont {Tatsumi}}]{Nakano:2003rd}%
  \BibitemOpen
  \bibfield  {author} {\bibinfo {author} {\bibfnamefont {E.}~\bibnamefont
  {Nakano}}, \bibinfo {author} {\bibfnamefont {T.}~\bibnamefont {Maruyama}}, \
  and\ \bibinfo {author} {\bibfnamefont {T.}~\bibnamefont {Tatsumi}},\ }\href
  {\doibase 10.1103/PhysRevD.68.105001} {\bibfield  {journal} {\bibinfo
  {journal} {Phys. Rev.}\ }\textbf {\bibinfo {volume} {D68}},\ \bibinfo {pages}
  {105001} (\bibinfo {year} {2003})},\ \Eprint
  {http://arxiv.org/abs/hep-ph/0304223} {arXiv:hep-ph/0304223 [hep-ph]}
  \BibitemShut {NoStop}%
\bibitem [{\citenamefont {Ohnishi}\ \emph {et~al.}(2007)\citenamefont
  {Ohnishi}, \citenamefont {Oka},\ and\ \citenamefont
  {Yasui}}]{Ohnishi:2006hs}%
  \BibitemOpen
  \bibfield  {author} {\bibinfo {author} {\bibfnamefont {K.}~\bibnamefont
  {Ohnishi}}, \bibinfo {author} {\bibfnamefont {M.}~\bibnamefont {Oka}}, \ and\
  \bibinfo {author} {\bibfnamefont {S.}~\bibnamefont {Yasui}},\ }\href
  {\doibase 10.1103/PhysRevD.76.097501} {\bibfield  {journal} {\bibinfo
  {journal} {Phys. Rev.}\ }\textbf {\bibinfo {volume} {D76}},\ \bibinfo {pages}
  {097501} (\bibinfo {year} {2007})},\ \Eprint
  {http://arxiv.org/abs/hep-ph/0609060} {arXiv:hep-ph/0609060 [hep-ph]}
  \BibitemShut {NoStop}%
\bibitem [{\citenamefont {Chamel}(2017)}]{Chamel2017}%
  \BibitemOpen
  \bibfield  {author} {\bibinfo {author} {\bibfnamefont {N.}~\bibnamefont
  {Chamel}},\ }\href {\doibase 10.1007/s12036-017-9470-9} {\bibfield  {journal}
  {\bibinfo  {journal} {Journal of Astrophysics and Astronomy}\ }\textbf
  {\bibinfo {volume} {38}},\ \bibinfo {pages} {43} (\bibinfo {year}
  {2017})}\BibitemShut {NoStop}%
\bibitem [{\citenamefont {Haskell}\ and\ \citenamefont
  {Sedrakian}(2018)}]{Haskell:2017lkl}%
  \BibitemOpen
  \bibfield  {author} {\bibinfo {author} {\bibfnamefont {B.}~\bibnamefont
  {Haskell}}\ and\ \bibinfo {author} {\bibfnamefont {A.}~\bibnamefont
  {Sedrakian}},\ }\href {\doibase 10.1007/978-3-319-97616-7_8} {\bibfield
  {journal} {\bibinfo  {journal} {Astrophys. Space Sci. Libr.}\ }\textbf
  {\bibinfo {volume} {457}},\ \bibinfo {pages} {401} (\bibinfo {year}
  {2018})},\ \Eprint {http://arxiv.org/abs/1709.10340} {arXiv:1709.10340
  [astro-ph.HE]} \BibitemShut {NoStop}%
\bibitem [{\citenamefont {Reichley}\ and\ \citenamefont
  {Downs}(1969)}]{Reichely1969}%
  \BibitemOpen
  \bibfield  {author} {\bibinfo {author} {\bibfnamefont {P.~E.}\ \bibnamefont
  {Reichley}}\ and\ \bibinfo {author} {\bibfnamefont {G.~S.}\ \bibnamefont
  {Downs}},\ }\href {\doibase 10.1038/222229a0} {\bibfield  {journal} {\bibinfo
   {journal} {Nature}\ }\textbf {\bibinfo {volume} {222}},\ \bibinfo {pages}
  {229} (\bibinfo {year} {1969})}\BibitemShut {NoStop}%
\bibitem [{\citenamefont {Baym}\ \emph {et~al.}(1969)\citenamefont {Baym},
  \citenamefont {Pethick}, \citenamefont {Pines},\ and\ \citenamefont
  {Ruderman}}]{Baym1969}%
  \BibitemOpen
  \bibfield  {author} {\bibinfo {author} {\bibfnamefont {G.}~\bibnamefont
  {Baym}}, \bibinfo {author} {\bibfnamefont {C.}~\bibnamefont {Pethick}},
  \bibinfo {author} {\bibfnamefont {D.}~\bibnamefont {Pines}}, \ and\ \bibinfo
  {author} {\bibfnamefont {M.}~\bibnamefont {Ruderman}},\ }\href {\doibase
  10.1038/224872a0} {\bibfield  {journal} {\bibinfo  {journal} {Nature}\
  }\textbf {\bibinfo {volume} {224}},\ \bibinfo {pages} {872} (\bibinfo {year}
  {1969})}\BibitemShut {NoStop}%
\bibitem [{\citenamefont {Pines}\ \emph {et~al.}(1972)\citenamefont {Pines},
  \citenamefont {Shaham},\ and\ \citenamefont {Ruderman}}]{Pines1972}%
  \BibitemOpen
  \bibfield  {author} {\bibinfo {author} {\bibfnamefont {D.}~\bibnamefont
  {Pines}}, \bibinfo {author} {\bibfnamefont {J.}~\bibnamefont {Shaham}}, \
  and\ \bibinfo {author} {\bibfnamefont {M.}~\bibnamefont {Ruderman}},\ }\href
  {\doibase 10.1038/physci237083a0} {\bibfield  {journal} {\bibinfo  {journal}
  {Nature Phys. Sci.}\ }\textbf {\bibinfo {volume} {237}},\ \bibinfo {pages}
  {83} (\bibinfo {year} {1972})}\BibitemShut {NoStop}%
\bibitem [{\citenamefont {Takatsuka}\ and\ \citenamefont
  {Tamagaki}(1988)}]{Takatsuka:1988kx}%
  \BibitemOpen
  \bibfield  {author} {\bibinfo {author} {\bibfnamefont {T.}~\bibnamefont
  {Takatsuka}}\ and\ \bibinfo {author} {\bibfnamefont {R.}~\bibnamefont
  {Tamagaki}},\ }\href {\doibase 10.1143/PTP.79.274} {\bibfield  {journal}
  {\bibinfo  {journal} {Prog. Theor. Phys.}\ }\textbf {\bibinfo {volume}
  {79}},\ \bibinfo {pages} {274} (\bibinfo {year} {1988})}\BibitemShut
  {NoStop}%
\bibitem [{\citenamefont {Anderson}\ and\ \citenamefont
  {Itoh}(1975)}]{Anderson1975}%
  \BibitemOpen
  \bibfield  {author} {\bibinfo {author} {\bibfnamefont {P.~W.}\ \bibnamefont
  {Anderson}}\ and\ \bibinfo {author} {\bibfnamefont {N.}~\bibnamefont
  {Itoh}},\ }\href {\doibase 10.1038/256025a0} {\bibfield  {journal} {\bibinfo
  {journal} {Nature}\ }\textbf {\bibinfo {volume} {256}},\ \bibinfo {pages}
  {25} (\bibinfo {year} {1975})}\BibitemShut {NoStop}%
\bibitem [{\citenamefont {Yakovlev}\ \emph {et~al.}(1999)\citenamefont
  {Yakovlev}, \citenamefont {Levenfish},\ and\ \citenamefont
  {Shibanov}}]{Yakovlev:1999sk}%
  \BibitemOpen
  \bibfield  {author} {\bibinfo {author} {\bibfnamefont {D.~G.}\ \bibnamefont
  {Yakovlev}}, \bibinfo {author} {\bibfnamefont {K.~P.}\ \bibnamefont
  {Levenfish}}, \ and\ \bibinfo {author} {\bibfnamefont {{\relax Yu}.~A.}\
  \bibnamefont {Shibanov}},\ }\href {\doibase 10.1070/PU1999v042n08ABEH000556}
  {\bibfield  {journal} {\bibinfo  {journal} {Phys. Usp.}\ }\textbf {\bibinfo
  {volume} {42}},\ \bibinfo {pages} {737} (\bibinfo {year} {1999})},\ \Eprint
  {http://arxiv.org/abs/astro-ph/9906456} {arXiv:astro-ph/9906456 [astro-ph]}
  \BibitemShut {NoStop}%
\bibitem [{\citenamefont {Heinke}\ and\ \citenamefont {Ho}(2010)}]{Heinke2010}%
  \BibitemOpen
  \bibfield  {author} {\bibinfo {author} {\bibfnamefont {C.~O.}\ \bibnamefont
  {Heinke}}\ and\ \bibinfo {author} {\bibfnamefont {W.~C.~G.}\ \bibnamefont
  {Ho}},\ }\href {http://stacks.iop.org/2041-8205/719/i=2/a=L167} {\bibfield
  {journal} {\bibinfo  {journal} {The Astrophysical Journal Letters}\ }\textbf
  {\bibinfo {volume} {719}},\ \bibinfo {pages} {L167} (\bibinfo {year}
  {2010})}\BibitemShut {NoStop}%
\bibitem [{\citenamefont {Shternin}\ \emph {et~al.}(2011)\citenamefont
  {Shternin}, \citenamefont {Yakovlev}, \citenamefont {Heinke}, \citenamefont
  {Ho},\ and\ \citenamefont {Patnaude}}]{Shternin2011}%
  \BibitemOpen
  \bibfield  {author} {\bibinfo {author} {\bibfnamefont {P.~S.}\ \bibnamefont
  {Shternin}}, \bibinfo {author} {\bibfnamefont {D.~G.}\ \bibnamefont
  {Yakovlev}}, \bibinfo {author} {\bibfnamefont {C.~O.}\ \bibnamefont
  {Heinke}}, \bibinfo {author} {\bibfnamefont {W.~C.~G.}\ \bibnamefont {Ho}}, \
  and\ \bibinfo {author} {\bibfnamefont {D.~J.}\ \bibnamefont {Patnaude}},\
  }\href {\doibase 10.1111/j.1745-3933.2011.01015.x} {\bibfield  {journal}
  {\bibinfo  {journal} {Monthly Notices of the Royal Astronomical Society:
  Letters}\ }\textbf {\bibinfo {volume} {412}},\ \bibinfo {pages} {L108}
  (\bibinfo {year} {2011})}\BibitemShut {NoStop}%
\bibitem [{\citenamefont {Page}\ \emph {et~al.}(2011)\citenamefont {Page},
  \citenamefont {Prakash}, \citenamefont {Lattimer},\ and\ \citenamefont
  {Steiner}}]{Page:2010aw}%
  \BibitemOpen
  \bibfield  {author} {\bibinfo {author} {\bibfnamefont {D.}~\bibnamefont
  {Page}}, \bibinfo {author} {\bibfnamefont {M.}~\bibnamefont {Prakash}},
  \bibinfo {author} {\bibfnamefont {J.~M.}\ \bibnamefont {Lattimer}}, \ and\
  \bibinfo {author} {\bibfnamefont {A.~W.}\ \bibnamefont {Steiner}},\ }\href
  {\doibase 10.1103/PhysRevLett.106.081101} {\bibfield  {journal} {\bibinfo
  {journal} {Phys. Rev. Lett.}\ }\textbf {\bibinfo {volume} {106}},\ \bibinfo
  {pages} {081101} (\bibinfo {year} {2011})},\ \Eprint
  {http://arxiv.org/abs/1011.6142} {arXiv:1011.6142 [astro-ph.HE]} \BibitemShut
  {NoStop}%
\bibitem [{\citenamefont {Dean}\ and\ \citenamefont
  {Hjorth-Jensen}(2003)}]{Dean:2002zx}%
  \BibitemOpen
  \bibfield  {author} {\bibinfo {author} {\bibfnamefont {D.~J.}\ \bibnamefont
  {Dean}}\ and\ \bibinfo {author} {\bibfnamefont {M.}~\bibnamefont
  {Hjorth-Jensen}},\ }\href {\doibase 10.1103/RevModPhys.75.607} {\bibfield
  {journal} {\bibinfo  {journal} {Rev. Mod. Phys.}\ }\textbf {\bibinfo {volume}
  {75}},\ \bibinfo {pages} {607} (\bibinfo {year} {2003})},\ \Eprint
  {http://arxiv.org/abs/nucl-th/0210033} {arXiv:nucl-th/0210033 [nucl-th]}
  \BibitemShut {NoStop}%
\bibitem [{\citenamefont {Migdal}(1960)}]{Migdal:1960}%
  \BibitemOpen
  \bibfield  {author} {\bibinfo {author} {\bibfnamefont {A.~B.}\ \bibnamefont
  {Migdal}},\ }\href@noop {} {\bibfield  {journal} {\bibinfo  {journal} {Zh.
  Eksp. Teor. Fiz.}\ }\textbf {\bibinfo {volume} {37}},\ \bibinfo {pages} {249}
  (\bibinfo {year} {1960})},\ \bibinfo {note} {[Sov. Phys.
  JETP10,no.1,176(1960)]}\BibitemShut {NoStop}%
\bibitem [{\citenamefont {{Wolf}}(1966)}]{1966ApJ...145..834W}%
  \BibitemOpen
  \bibfield  {author} {\bibinfo {author} {\bibfnamefont {R.~A.}\ \bibnamefont
  {{Wolf}}},\ }\href {\doibase 10.1086/148829} {\bibfield  {journal} {\bibinfo
  {journal} {\apj}\ }\textbf {\bibinfo {volume} {145}},\ \bibinfo {pages} {834}
  (\bibinfo {year} {1966})}\BibitemShut {NoStop}%
\bibitem [{\citenamefont {Tabakin}(1968)}]{Tabakin:1968zz}%
  \BibitemOpen
  \bibfield  {author} {\bibinfo {author} {\bibfnamefont {F.}~\bibnamefont
  {Tabakin}},\ }\href {\doibase 10.1103/PhysRev.174.1208} {\bibfield  {journal}
  {\bibinfo  {journal} {Phys. Rev.}\ }\textbf {\bibinfo {volume} {174}},\
  \bibinfo {pages} {1208} (\bibinfo {year} {1968})}\BibitemShut {NoStop}%
\bibitem [{\citenamefont {Hoffberg}\ \emph {et~al.}(1970)\citenamefont
  {Hoffberg}, \citenamefont {Glassgold}, \citenamefont {Richardson},\ and\
  \citenamefont {Ruderman}}]{Hoffberg:1970vqj}%
  \BibitemOpen
  \bibfield  {author} {\bibinfo {author} {\bibfnamefont {M.}~\bibnamefont
  {Hoffberg}}, \bibinfo {author} {\bibfnamefont {A.~E.}\ \bibnamefont
  {Glassgold}}, \bibinfo {author} {\bibfnamefont {R.~W.}\ \bibnamefont
  {Richardson}}, \ and\ \bibinfo {author} {\bibfnamefont {M.}~\bibnamefont
  {Ruderman}},\ }\href {\doibase 10.1103/PhysRevLett.24.775} {\bibfield
  {journal} {\bibinfo  {journal} {Phys. Rev. Lett.}\ }\textbf {\bibinfo
  {volume} {24}},\ \bibinfo {pages} {775} (\bibinfo {year} {1970})}\BibitemShut
  {NoStop}%
\bibitem [{\citenamefont {Tamagaki}(1970)}]{Tamagaki1970}%
  \BibitemOpen
  \bibfield  {author} {\bibinfo {author} {\bibfnamefont {R.}~\bibnamefont
  {Tamagaki}},\ }\href {\doibase 10.1143/PTP.44.905} {\bibfield  {journal}
  {\bibinfo  {journal} {Progress of Theoretical Physics}\ }\textbf {\bibinfo
  {volume} {44}},\ \bibinfo {pages} {905} (\bibinfo {year} {1970})}\BibitemShut
  {NoStop}%
\bibitem [{\citenamefont {Takatsuka}\ and\ \citenamefont
  {Tamagaki}(1971)}]{Takatsuka1971}%
  \BibitemOpen
  \bibfield  {author} {\bibinfo {author} {\bibfnamefont {T.}~\bibnamefont
  {Takatsuka}}\ and\ \bibinfo {author} {\bibfnamefont {R.}~\bibnamefont
  {Tamagaki}},\ }\href {\doibase 10.1143/PTP.46.114} {\bibfield  {journal}
  {\bibinfo  {journal} {Progress of Theoretical Physics}\ }\textbf {\bibinfo
  {volume} {46}},\ \bibinfo {pages} {114} (\bibinfo {year} {1971})}\BibitemShut
  {NoStop}%
\bibitem [{\citenamefont {Takatsuka}(1972)}]{Takatsuka1972}%
  \BibitemOpen
  \bibfield  {author} {\bibinfo {author} {\bibfnamefont {T.}~\bibnamefont
  {Takatsuka}},\ }\href {\doibase 10.1143/PTP.47.1062} {\bibfield  {journal}
  {\bibinfo  {journal} {Progress of Theoretical Physics}\ }\textbf {\bibinfo
  {volume} {47}},\ \bibinfo {pages} {1062} (\bibinfo {year}
  {1972})}\BibitemShut {NoStop}%
\bibitem [{\citenamefont {Takatsuka}\ and\ \citenamefont
  {Tamagaki}(1993)}]{Takatsuka:1992ga}%
  \BibitemOpen
  \bibfield  {author} {\bibinfo {author} {\bibfnamefont {T.}~\bibnamefont
  {Takatsuka}}\ and\ \bibinfo {author} {\bibfnamefont {R.}~\bibnamefont
  {Tamagaki}},\ }\href {\doibase 10.1143/PTPS.112.27} {\bibfield  {journal}
  {\bibinfo  {journal} {Prog. Theor. Phys. Suppl.}\ }\textbf {\bibinfo {volume}
  {112}},\ \bibinfo {pages} {27} (\bibinfo {year} {1993})}\BibitemShut
  {NoStop}%
\bibitem [{\citenamefont {Baldo}\ \emph {et~al.}(1992)\citenamefont {Baldo},
  \citenamefont {Cugnon}, \citenamefont {Lejeune},\ and\ \citenamefont
  {Lombardo}}]{Baldo:1992kzz}%
  \BibitemOpen
  \bibfield  {author} {\bibinfo {author} {\bibfnamefont {M.}~\bibnamefont
  {Baldo}}, \bibinfo {author} {\bibfnamefont {J.}~\bibnamefont {Cugnon}},
  \bibinfo {author} {\bibfnamefont {A.}~\bibnamefont {Lejeune}}, \ and\
  \bibinfo {author} {\bibfnamefont {U.}~\bibnamefont {Lombardo}},\ }\href
  {\doibase 10.1016/0375-9474(92)90387-Y} {\bibfield  {journal} {\bibinfo
  {journal} {Nucl. Phys.}\ }\textbf {\bibinfo {volume} {A536}},\ \bibinfo
  {pages} {349} (\bibinfo {year} {1992})}\BibitemShut {NoStop}%
\bibitem [{\citenamefont {Elgaroy}\ \emph {et~al.}(1996)\citenamefont
  {Elgaroy}, \citenamefont {Engvik}, \citenamefont {Hjorth-Jensen},\ and\
  \citenamefont {Osnes}}]{Elgaroy:1996hp}%
  \BibitemOpen
  \bibfield  {author} {\bibinfo {author} {\bibfnamefont {O.}~\bibnamefont
  {Elgaroy}}, \bibinfo {author} {\bibfnamefont {L.}~\bibnamefont {Engvik}},
  \bibinfo {author} {\bibfnamefont {M.}~\bibnamefont {Hjorth-Jensen}}, \ and\
  \bibinfo {author} {\bibfnamefont {E.}~\bibnamefont {Osnes}},\ }\href
  {\doibase 10.1016/0375-9474(96)00217-5} {\bibfield  {journal} {\bibinfo
  {journal} {Nucl. Phys.}\ }\textbf {\bibinfo {volume} {A607}},\ \bibinfo
  {pages} {425} (\bibinfo {year} {1996})},\ \Eprint
  {http://arxiv.org/abs/nucl-th/9604032} {arXiv:nucl-th/9604032 [nucl-th]}
  \BibitemShut {NoStop}%
\bibitem [{\citenamefont {Khodel}\ \emph {et~al.}(1998)\citenamefont {Khodel},
  \citenamefont {Khodel},\ and\ \citenamefont {Clark}}]{Khodel:1998hn}%
  \BibitemOpen
  \bibfield  {author} {\bibinfo {author} {\bibfnamefont {V.~A.}\ \bibnamefont
  {Khodel}}, \bibinfo {author} {\bibfnamefont {V.~V.}\ \bibnamefont {Khodel}},
  \ and\ \bibinfo {author} {\bibfnamefont {J.~W.}\ \bibnamefont {Clark}},\
  }\href {\doibase 10.1103/PhysRevLett.81.3828} {\bibfield  {journal} {\bibinfo
   {journal} {Phys. Rev. Lett.}\ }\textbf {\bibinfo {volume} {81}},\ \bibinfo
  {pages} {3828} (\bibinfo {year} {1998})},\ \Eprint
  {http://arxiv.org/abs/nucl-th/9807034} {arXiv:nucl-th/9807034 [nucl-th]}
  \BibitemShut {NoStop}%
\bibitem [{\citenamefont {Baldo}\ \emph {et~al.}(1998)\citenamefont {Baldo},
  \citenamefont {Elgaroey}, \citenamefont {Engvik}, \citenamefont
  {Hjorth-Jensen},\ and\ \citenamefont {Schulze}}]{Baldo:1998ca}%
  \BibitemOpen
  \bibfield  {author} {\bibinfo {author} {\bibfnamefont {M.}~\bibnamefont
  {Baldo}}, \bibinfo {author} {\bibfnamefont {O.}~\bibnamefont {Elgaroey}},
  \bibinfo {author} {\bibfnamefont {L.}~\bibnamefont {Engvik}}, \bibinfo
  {author} {\bibfnamefont {M.}~\bibnamefont {Hjorth-Jensen}}, \ and\ \bibinfo
  {author} {\bibfnamefont {H.~J.}\ \bibnamefont {Schulze}},\ }\href {\doibase
  10.1103/PhysRevC.58.1921} {\bibfield  {journal} {\bibinfo  {journal} {Phys.
  Rev.}\ }\textbf {\bibinfo {volume} {C58}},\ \bibinfo {pages} {1921} (\bibinfo
  {year} {1998})},\ \Eprint {http://arxiv.org/abs/nucl-th/9806097}
  {arXiv:nucl-th/9806097 [nucl-th]} \BibitemShut {NoStop}%
\bibitem [{\citenamefont {Khodel}\ \emph {et~al.}(2001)\citenamefont {Khodel},
  \citenamefont {Khodel},\ and\ \citenamefont {Clark}}]{Khodel:2000qw}%
  \BibitemOpen
  \bibfield  {author} {\bibinfo {author} {\bibfnamefont {V.~V.}\ \bibnamefont
  {Khodel}}, \bibinfo {author} {\bibfnamefont {V.~A.}\ \bibnamefont {Khodel}},
  \ and\ \bibinfo {author} {\bibfnamefont {J.~W.}\ \bibnamefont {Clark}},\
  }\href {\doibase 10.1016/S0375-9474(00)00351-1} {\bibfield  {journal}
  {\bibinfo  {journal} {Nucl. Phys.}\ }\textbf {\bibinfo {volume} {A679}},\
  \bibinfo {pages} {827} (\bibinfo {year} {2001})},\ \Eprint
  {http://arxiv.org/abs/nucl-th/0001006} {arXiv:nucl-th/0001006 [nucl-th]}
  \BibitemShut {NoStop}%
\bibitem [{\citenamefont {Zverev}\ \emph {et~al.}(2003)\citenamefont {Zverev},
  \citenamefont {Clark},\ and\ \citenamefont {Khodel}}]{Zverev:2003ak}%
  \BibitemOpen
  \bibfield  {author} {\bibinfo {author} {\bibfnamefont {M.~V.}\ \bibnamefont
  {Zverev}}, \bibinfo {author} {\bibfnamefont {J.~W.}\ \bibnamefont {Clark}}, \
  and\ \bibinfo {author} {\bibfnamefont {V.~A.}\ \bibnamefont {Khodel}},\
  }\href {\doibase 10.1016/S0375-9474(03)00653-5} {\bibfield  {journal}
  {\bibinfo  {journal} {Nucl. Phys.}\ }\textbf {\bibinfo {volume} {A720}},\
  \bibinfo {pages} {20} (\bibinfo {year} {2003})},\ \Eprint
  {http://arxiv.org/abs/nucl-th/0301028} {arXiv:nucl-th/0301028 [nucl-th]}
  \BibitemShut {NoStop}%
\bibitem [{\citenamefont {Maurizio}\ \emph {et~al.}(2014)\citenamefont
  {Maurizio}, \citenamefont {Holt},\ and\ \citenamefont
  {Finelli}}]{Maurizio:2014qsa}%
  \BibitemOpen
  \bibfield  {author} {\bibinfo {author} {\bibfnamefont {S.}~\bibnamefont
  {Maurizio}}, \bibinfo {author} {\bibfnamefont {J.~W.}\ \bibnamefont {Holt}},
  \ and\ \bibinfo {author} {\bibfnamefont {P.}~\bibnamefont {Finelli}},\ }\href
  {\doibase 10.1103/PhysRevC.90.044003} {\bibfield  {journal} {\bibinfo
  {journal} {Phys. Rev.}\ }\textbf {\bibinfo {volume} {C90}},\ \bibinfo {pages}
  {044003} (\bibinfo {year} {2014})},\ \Eprint {http://arxiv.org/abs/1408.6281}
  {arXiv:1408.6281 [nucl-th]} \BibitemShut {NoStop}%
\bibitem [{\citenamefont {Bogner}\ \emph {et~al.}(2010)\citenamefont {Bogner},
  \citenamefont {Furnstahl},\ and\ \citenamefont {Schwenk}}]{Bogner:2009bt}%
  \BibitemOpen
  \bibfield  {author} {\bibinfo {author} {\bibfnamefont {S.~K.}\ \bibnamefont
  {Bogner}}, \bibinfo {author} {\bibfnamefont {R.~J.}\ \bibnamefont
  {Furnstahl}}, \ and\ \bibinfo {author} {\bibfnamefont {A.}~\bibnamefont
  {Schwenk}},\ }\href {\doibase 10.1016/j.ppnp.2010.03.001} {\bibfield
  {journal} {\bibinfo  {journal} {Prog. Part. Nucl. Phys.}\ }\textbf {\bibinfo
  {volume} {65}},\ \bibinfo {pages} {94} (\bibinfo {year} {2010})},\ \Eprint
  {http://arxiv.org/abs/0912.3688} {arXiv:0912.3688 [nucl-th]} \BibitemShut
  {NoStop}%
\bibitem [{\citenamefont {Srinivas}\ and\ \citenamefont
  {Ramanan}(2016)}]{Srinivas:2016kir}%
  \BibitemOpen
  \bibfield  {author} {\bibinfo {author} {\bibfnamefont {S.}~\bibnamefont
  {Srinivas}}\ and\ \bibinfo {author} {\bibfnamefont {S.}~\bibnamefont
  {Ramanan}},\ }\href {\doibase 10.1103/PhysRevC.94.064303} {\bibfield
  {journal} {\bibinfo  {journal} {Phys. Rev.}\ }\textbf {\bibinfo {volume}
  {C94}},\ \bibinfo {pages} {064303} (\bibinfo {year} {2016})},\ \Eprint
  {http://arxiv.org/abs/1606.09053} {arXiv:1606.09053 [nucl-th]} \BibitemShut
  {NoStop}%
\bibitem [{\citenamefont {Bedaque}\ \emph {et~al.}(2003)\citenamefont
  {Bedaque}, \citenamefont {Rupak},\ and\ \citenamefont
  {Savage}}]{Bedaque:2003wj}%
  \BibitemOpen
  \bibfield  {author} {\bibinfo {author} {\bibfnamefont {P.~F.}\ \bibnamefont
  {Bedaque}}, \bibinfo {author} {\bibfnamefont {G.}~\bibnamefont {Rupak}}, \
  and\ \bibinfo {author} {\bibfnamefont {M.~J.}\ \bibnamefont {Savage}},\
  }\href {\doibase 10.1103/PhysRevC.68.065802} {\bibfield  {journal} {\bibinfo
  {journal} {Phys. Rev.}\ }\textbf {\bibinfo {volume} {C68}},\ \bibinfo {pages}
  {065802} (\bibinfo {year} {2003})},\ \Eprint
  {http://arxiv.org/abs/nucl-th/0305032} {arXiv:nucl-th/0305032 [nucl-th]}
  \BibitemShut {NoStop}%
\bibitem [{\citenamefont {Bedaque}\ and\ \citenamefont
  {Nicholson}(2013)}]{Bedaque:2012bs}%
  \BibitemOpen
  \bibfield  {author} {\bibinfo {author} {\bibfnamefont {P.~F.}\ \bibnamefont
  {Bedaque}}\ and\ \bibinfo {author} {\bibfnamefont {A.~N.}\ \bibnamefont
  {Nicholson}},\ }\href {\doibase 10.1103/PhysRevC.89.029902,
  10.1103/PhysRevC.87.055807} {\bibfield  {journal} {\bibinfo  {journal} {Phys.
  Rev.}\ }\textbf {\bibinfo {volume} {C87}},\ \bibinfo {pages} {055807}
  (\bibinfo {year} {2013})},\ \bibinfo {note} {[Erratum: Phys.
  Rev.C89,no.2,029902(2014)]},\ \Eprint {http://arxiv.org/abs/1212.1122}
  {arXiv:1212.1122 [nucl-th]} \BibitemShut {NoStop}%
\bibitem [{\citenamefont {Bedaque}\ and\ \citenamefont
  {Reddy}(2014)}]{Bedaque:2013fja}%
  \BibitemOpen
  \bibfield  {author} {\bibinfo {author} {\bibfnamefont {P.~F.}\ \bibnamefont
  {Bedaque}}\ and\ \bibinfo {author} {\bibfnamefont {S.}~\bibnamefont
  {Reddy}},\ }\href {\doibase 10.1016/j.physletb.2014.06.033} {\bibfield
  {journal} {\bibinfo  {journal} {Phys. Lett.}\ }\textbf {\bibinfo {volume}
  {B735}},\ \bibinfo {pages} {340} (\bibinfo {year} {2014})},\ \Eprint
  {http://arxiv.org/abs/1307.8183} {arXiv:1307.8183 [nucl-th]} \BibitemShut
  {NoStop}%
\bibitem [{\citenamefont {Bedaque}\ \emph {et~al.}(2015)\citenamefont
  {Bedaque}, \citenamefont {Nicholson},\ and\ \citenamefont
  {Sen}}]{Bedaque:2014zta}%
  \BibitemOpen
  \bibfield  {author} {\bibinfo {author} {\bibfnamefont {P.~F.}\ \bibnamefont
  {Bedaque}}, \bibinfo {author} {\bibfnamefont {A.~N.}\ \bibnamefont
  {Nicholson}}, \ and\ \bibinfo {author} {\bibfnamefont {S.}~\bibnamefont
  {Sen}},\ }\href {\doibase 10.1103/PhysRevC.92.035809} {\bibfield  {journal}
  {\bibinfo  {journal} {Phys. Rev.}\ }\textbf {\bibinfo {volume} {C92}},\
  \bibinfo {pages} {035809} (\bibinfo {year} {2015})},\ \Eprint
  {http://arxiv.org/abs/1408.5145} {arXiv:1408.5145 [nucl-th]} \BibitemShut
  {NoStop}%
\bibitem [{\citenamefont {Leinson}(2010{\natexlab{a}})}]{Leinson:2009nu}%
  \BibitemOpen
  \bibfield  {author} {\bibinfo {author} {\bibfnamefont {L.~B.}\ \bibnamefont
  {Leinson}},\ }\href {\doibase 10.1103/PhysRevC.81.025501} {\bibfield
  {journal} {\bibinfo  {journal} {Phys. Rev.}\ }\textbf {\bibinfo {volume}
  {C81}},\ \bibinfo {pages} {025501} (\bibinfo {year} {2010}{\natexlab{a}})},\
  \Eprint {http://arxiv.org/abs/0912.2164} {arXiv:0912.2164 [astro-ph.SR]}
  \BibitemShut {NoStop}%
\bibitem [{\citenamefont {Leinson}(2010{\natexlab{b}})}]{Leinson:2010yf}%
  \BibitemOpen
  \bibfield  {author} {\bibinfo {author} {\bibfnamefont {L.~B.}\ \bibnamefont
  {Leinson}},\ }\href {\doibase 10.1016/j.physletb.2010.04.046} {\bibfield
  {journal} {\bibinfo  {journal} {Phys. Lett.}\ }\textbf {\bibinfo {volume}
  {B689}},\ \bibinfo {pages} {60} (\bibinfo {year} {2010}{\natexlab{b}})},\
  \Eprint {http://arxiv.org/abs/1001.2617} {arXiv:1001.2617 [astro-ph.SR]}
  \BibitemShut {NoStop}%
\bibitem [{\citenamefont {Leinson}(2010{\natexlab{c}})}]{Leinson:2010pk}%
  \BibitemOpen
  \bibfield  {author} {\bibinfo {author} {\bibfnamefont {L.~B.}\ \bibnamefont
  {Leinson}},\ }\href {\doibase 10.1103/PhysRevC.82.065503,
  10.1103/PhysRevC.84.049901} {\bibfield  {journal} {\bibinfo  {journal} {Phys.
  Rev.}\ }\textbf {\bibinfo {volume} {C82}},\ \bibinfo {pages} {065503}
  (\bibinfo {year} {2010}{\natexlab{c}})},\ \bibinfo {note} {[Erratum: Phys.
  Rev.C84,049901(2011)]},\ \Eprint {http://arxiv.org/abs/1012.5387}
  {arXiv:1012.5387 [hep-ph]} \BibitemShut {NoStop}%
\bibitem [{\citenamefont {Leinson}(2011{\natexlab{a}})}]{Leinson:2010ru}%
  \BibitemOpen
  \bibfield  {author} {\bibinfo {author} {\bibfnamefont {L.~B.}\ \bibnamefont
  {Leinson}},\ }\href {\doibase 10.1103/PhysRevC.83.055803} {\bibfield
  {journal} {\bibinfo  {journal} {Phys. Rev.}\ }\textbf {\bibinfo {volume}
  {C83}},\ \bibinfo {pages} {055803} (\bibinfo {year} {2011}{\natexlab{a}})},\
  \Eprint {http://arxiv.org/abs/1007.2803} {arXiv:1007.2803 [hep-ph]}
  \BibitemShut {NoStop}%
\bibitem [{\citenamefont {Leinson}(2011{\natexlab{b}})}]{Leinson:2011jr}%
  \BibitemOpen
  \bibfield  {author} {\bibinfo {author} {\bibfnamefont {L.~B.}\ \bibnamefont
  {Leinson}},\ }\href {\doibase 10.1103/PhysRevC.84.045501} {\bibfield
  {journal} {\bibinfo  {journal} {Phys. Rev.}\ }\textbf {\bibinfo {volume}
  {C84}},\ \bibinfo {pages} {045501} (\bibinfo {year} {2011}{\natexlab{b}})},\
  \Eprint {http://arxiv.org/abs/1110.2145} {arXiv:1110.2145 [nucl-th]}
  \BibitemShut {NoStop}%
\bibitem [{\citenamefont {Leinson}(2012)}]{Leinson:2012pn}%
  \BibitemOpen
  \bibfield  {author} {\bibinfo {author} {\bibfnamefont {L.~B.}\ \bibnamefont
  {Leinson}},\ }\href {\doibase 10.1103/PhysRevC.85.065502} {\bibfield
  {journal} {\bibinfo  {journal} {Phys. Rev.}\ }\textbf {\bibinfo {volume}
  {C85}},\ \bibinfo {pages} {065502} (\bibinfo {year} {2012})},\ \Eprint
  {http://arxiv.org/abs/1206.3648} {arXiv:1206.3648 [nucl-th]} \BibitemShut
  {NoStop}%
\bibitem [{\citenamefont {Leinson}(2013)}]{Leinson:2013si}%
  \BibitemOpen
  \bibfield  {author} {\bibinfo {author} {\bibfnamefont {L.~B.}\ \bibnamefont
  {Leinson}},\ }\href {\doibase 10.1103/PhysRevC.87.025501} {\bibfield
  {journal} {\bibinfo  {journal} {Phys. Rev.}\ }\textbf {\bibinfo {volume}
  {C87}},\ \bibinfo {pages} {025501} (\bibinfo {year} {2013})},\ \Eprint
  {http://arxiv.org/abs/1301.5439} {arXiv:1301.5439 [nucl-th]} \BibitemShut
  {NoStop}%
\bibitem [{\citenamefont {Leinson}(2015)}]{Leinson:2014cja}%
  \BibitemOpen
  \bibfield  {author} {\bibinfo {author} {\bibfnamefont {L.~B.}\ \bibnamefont
  {Leinson}},\ }\href {\doibase 10.1016/j.physletb.2014.12.017} {\bibfield
  {journal} {\bibinfo  {journal} {Phys. Lett.}\ }\textbf {\bibinfo {volume}
  {B741}},\ \bibinfo {pages} {87} (\bibinfo {year} {2015})},\ \Eprint
  {http://arxiv.org/abs/1411.6833} {arXiv:1411.6833 [astro-ph.SR]} \BibitemShut
  {NoStop}%
\bibitem [{\citenamefont {Shahabasyan}\ and\ \citenamefont
  {Shahabasyan}(2011)}]{Shahabasyan:2011zz}%
  \BibitemOpen
  \bibfield  {author} {\bibinfo {author} {\bibfnamefont {K.~M.}\ \bibnamefont
  {Shahabasyan}}\ and\ \bibinfo {author} {\bibfnamefont {M.~K.}\ \bibnamefont
  {Shahabasyan}},\ }\href {\doibase 10.1007/s10511-011-9193-6} {\bibfield
  {journal} {\bibinfo  {journal} {Astrophysics}\ }\textbf {\bibinfo {volume}
  {54}},\ \bibinfo {pages} {429} (\bibinfo {year} {2011})},\ \bibinfo {note}
  {[Astrofiz.54,483(2011)]}\BibitemShut {NoStop}%
\bibitem [{\citenamefont {Fujita}\ and\ \citenamefont
  {Tsuneto}(1972)}]{Fujita1972}%
  \BibitemOpen
  \bibfield  {author} {\bibinfo {author} {\bibfnamefont {T.}~\bibnamefont
  {Fujita}}\ and\ \bibinfo {author} {\bibfnamefont {T.}~\bibnamefont
  {Tsuneto}},\ }\href {\doibase 10.1143/PTP.48.766} {\bibfield  {journal}
  {\bibinfo  {journal} {Progress of Theoretical Physics}\ }\textbf {\bibinfo
  {volume} {48}},\ \bibinfo {pages} {766} (\bibinfo {year} {1972})}\BibitemShut
  {NoStop}%
\bibitem [{\citenamefont {Richardson}(1972)}]{Richardson:1972xn}%
  \BibitemOpen
  \bibfield  {author} {\bibinfo {author} {\bibfnamefont {R.~W.}\ \bibnamefont
  {Richardson}},\ }\href {\doibase 10.1103/PhysRevD.5.1883} {\bibfield
  {journal} {\bibinfo  {journal} {Phys. Rev.}\ }\textbf {\bibinfo {volume}
  {D5}},\ \bibinfo {pages} {1883} (\bibinfo {year} {1972})}\BibitemShut
  {NoStop}%
\bibitem [{\citenamefont {Sauls}\ and\ \citenamefont
  {Serene}(1978)}]{Sauls:1978lna}%
  \BibitemOpen
  \bibfield  {author} {\bibinfo {author} {\bibfnamefont {J.~A.}\ \bibnamefont
  {Sauls}}\ and\ \bibinfo {author} {\bibfnamefont {J.}~\bibnamefont {Serene}},\
  }\href {\doibase 10.1103/PhysRevD.17.1524} {\bibfield  {journal} {\bibinfo
  {journal} {Phys. Rev.}\ }\textbf {\bibinfo {volume} {D17}},\ \bibinfo {pages}
  {1524} (\bibinfo {year} {1978})}\BibitemShut {NoStop}%
\bibitem [{\citenamefont {Muzikar}\ \emph {et~al.}(1980)\citenamefont
  {Muzikar}, \citenamefont {Sauls},\ and\ \citenamefont
  {Serene}}]{Muzikar:1980as}%
  \BibitemOpen
  \bibfield  {author} {\bibinfo {author} {\bibfnamefont {P.}~\bibnamefont
  {Muzikar}}, \bibinfo {author} {\bibfnamefont {J.~A.}\ \bibnamefont {Sauls}},
  \ and\ \bibinfo {author} {\bibfnamefont {J.~W.}\ \bibnamefont {Serene}},\
  }\href {\doibase 10.1103/PhysRevD.21.1494} {\bibfield  {journal} {\bibinfo
  {journal} {Phys. Rev.}\ }\textbf {\bibinfo {volume} {D21}},\ \bibinfo {pages}
  {1494} (\bibinfo {year} {1980})}\BibitemShut {NoStop}%
\bibitem [{\citenamefont {Sauls}\ \emph {et~al.}(1982)\citenamefont {Sauls},
  \citenamefont {Stein},\ and\ \citenamefont {Serene}}]{Sauls:1982ie}%
  \BibitemOpen
  \bibfield  {author} {\bibinfo {author} {\bibfnamefont {J.~A.}\ \bibnamefont
  {Sauls}}, \bibinfo {author} {\bibfnamefont {D.~L.}\ \bibnamefont {Stein}}, \
  and\ \bibinfo {author} {\bibfnamefont {J.~W.}\ \bibnamefont {Serene}},\
  }\href {\doibase 10.1103/PhysRevD.25.967} {\bibfield  {journal} {\bibinfo
  {journal} {Phys. Rev.}\ }\textbf {\bibinfo {volume} {D25}},\ \bibinfo {pages}
  {967} (\bibinfo {year} {1982})}\BibitemShut {NoStop}%
\bibitem [{\citenamefont {Vulovic}\ and\ \citenamefont
  {Sauls}(1984)}]{Vulovic:1984kc}%
  \BibitemOpen
  \bibfield  {author} {\bibinfo {author} {\bibfnamefont {V.~Z.}\ \bibnamefont
  {Vulovic}}\ and\ \bibinfo {author} {\bibfnamefont {J.~A.}\ \bibnamefont
  {Sauls}},\ }\href {\doibase 10.1103/PhysRevD.29.2705} {\bibfield  {journal}
  {\bibinfo  {journal} {Phys. Rev.}\ }\textbf {\bibinfo {volume} {D29}},\
  \bibinfo {pages} {2705} (\bibinfo {year} {1984})}\BibitemShut {NoStop}%
\bibitem [{\citenamefont {Masuda}\ and\ \citenamefont
  {Nitta}(2016)}]{Masuda:2015jka}%
  \BibitemOpen
  \bibfield  {author} {\bibinfo {author} {\bibfnamefont {K.}~\bibnamefont
  {Masuda}}\ and\ \bibinfo {author} {\bibfnamefont {M.}~\bibnamefont {Nitta}},\
  }\href {\doibase 10.1103/PhysRevC.93.035804} {\bibfield  {journal} {\bibinfo
  {journal} {Phys. Rev.}\ }\textbf {\bibinfo {volume} {C93}},\ \bibinfo {pages}
  {035804} (\bibinfo {year} {2016})},\ \Eprint
  {http://arxiv.org/abs/1512.01946} {arXiv:1512.01946 [nucl-th]} \BibitemShut
  {NoStop}%
\bibitem [{\citenamefont {Masuda}\ and\ \citenamefont
  {Nitta}()}]{Masuda:2016vak}%
  \BibitemOpen
  \bibfield  {author} {\bibinfo {author} {\bibfnamefont {K.}~\bibnamefont
  {Masuda}}\ and\ \bibinfo {author} {\bibfnamefont {M.}~\bibnamefont {Nitta}},\
  }\href@noop {} {\ }\Eprint {http://arxiv.org/abs/1602.07050}
  {arXiv:1602.07050 [nucl-th]} \BibitemShut {NoStop}%
\bibitem [{\citenamefont {Mermin}(1974)}]{Mermin:1974zz}%
  \BibitemOpen
  \bibfield  {author} {\bibinfo {author} {\bibfnamefont {N.~D.}\ \bibnamefont
  {Mermin}},\ }\href {\doibase 10.1103/PhysRevA.9.868} {\bibfield  {journal}
  {\bibinfo  {journal} {Phys. Rev.}\ }\textbf {\bibinfo {volume} {A9}},\
  \bibinfo {pages} {868} (\bibinfo {year} {1974})}\BibitemShut {NoStop}%
\bibitem [{\citenamefont {Vollhardt}\ and\ \citenamefont
  {W{\"o}lfle}(2013)}]{vollhardt2013superfluid}%
  \BibitemOpen
  \bibfield  {author} {\bibinfo {author} {\bibfnamefont {D.}~\bibnamefont
  {Vollhardt}}\ and\ \bibinfo {author} {\bibfnamefont {P.}~\bibnamefont
  {W{\"o}lfle}},\ }\href {https://books.google.co.jp/books?id=jY6yAAAAQBAJ}
  {\emph {\bibinfo {title} {The Superfluid Phases of Helium 3}}},\ Dover Books
  on Physics Series\ (\bibinfo  {publisher} {Dover Publications, New York},\
  \bibinfo {year} {2013})\BibitemShut {NoStop}%
\bibitem [{\citenamefont {Mackenzie}\ and\ \citenamefont
  {Maeno}(2003)}]{RevModPhys.75.657}%
  \BibitemOpen
  \bibfield  {author} {\bibinfo {author} {\bibfnamefont {A.~P.}\ \bibnamefont
  {Mackenzie}}\ and\ \bibinfo {author} {\bibfnamefont {Y.}~\bibnamefont
  {Maeno}},\ }\href {\doibase 10.1103/RevModPhys.75.657} {\bibfield  {journal}
  {\bibinfo  {journal} {Rev. Mod. Phys.}\ }\textbf {\bibinfo {volume} {75}},\
  \bibinfo {pages} {657} (\bibinfo {year} {2003})}\BibitemShut {NoStop}%
\bibitem [{\citenamefont {{Kawaguchi}}\ and\ \citenamefont
  {{Ueda}}(2012)}]{2010arXiv1001.2072K}%
  \BibitemOpen
  \bibfield  {author} {\bibinfo {author} {\bibfnamefont {Y.}~\bibnamefont
  {{Kawaguchi}}}\ and\ \bibinfo {author} {\bibfnamefont {M.}~\bibnamefont
  {{Ueda}}},\ }\href {\doibase 10.1016/j.physrep.2012.07.005} {\bibfield
  {journal} {\bibinfo  {journal} {Phys. Rep.}\ }\textbf {\bibinfo {volume}
  {520}},\ \bibinfo {pages} {253 } (\bibinfo {year} {2012})},\ \Eprint
  {http://arxiv.org/abs/1001.2072} {arXiv:1001.2072 [cond-mat.quant-gas]}
  \BibitemShut {NoStop}%
\bibitem [{\citenamefont {Mizushima}\ \emph {et~al.}(2017)\citenamefont
  {Mizushima}, \citenamefont {Masuda},\ and\ \citenamefont
  {Nitta}}]{Mizushima:2016fbn}%
  \BibitemOpen
  \bibfield  {author} {\bibinfo {author} {\bibfnamefont {T.}~\bibnamefont
  {Mizushima}}, \bibinfo {author} {\bibfnamefont {K.}~\bibnamefont {Masuda}}, \
  and\ \bibinfo {author} {\bibfnamefont {M.}~\bibnamefont {Nitta}},\ }\href
  {\doibase 10.1103/PhysRevB.95.140503} {\bibfield  {journal} {\bibinfo
  {journal} {Phys. Rev.}\ }\textbf {\bibinfo {volume} {B95}},\ \bibinfo {pages}
  {140503} (\bibinfo {year} {2017})},\ \Eprint
  {http://arxiv.org/abs/1607.07266} {arXiv:1607.07266 [cond-mat.supr-con]}
  \BibitemShut {NoStop}%
\bibitem [{\citenamefont {Chatterjee}\ \emph {et~al.}(2017)\citenamefont
  {Chatterjee}, \citenamefont {Haberichter},\ and\ \citenamefont
  {Nitta}}]{Chatterjee:2016gpm}%
  \BibitemOpen
  \bibfield  {author} {\bibinfo {author} {\bibfnamefont {C.}~\bibnamefont
  {Chatterjee}}, \bibinfo {author} {\bibfnamefont {M.}~\bibnamefont
  {Haberichter}}, \ and\ \bibinfo {author} {\bibfnamefont {M.}~\bibnamefont
  {Nitta}},\ }\href {\doibase 10.1103/PhysRevC.96.055807} {\bibfield  {journal}
  {\bibinfo  {journal} {Phys. Rev.}\ }\textbf {\bibinfo {volume} {C96}},\
  \bibinfo {pages} {055807} (\bibinfo {year} {2017})},\ \Eprint
  {http://arxiv.org/abs/1612.05588} {arXiv:1612.05588 [nucl-th]} \BibitemShut
  {NoStop}%
\bibitem [{\citenamefont {Uchino}\ \emph {et~al.}(2010)\citenamefont {Uchino},
  \citenamefont {Kobayashi}, \citenamefont {Nitta},\ and\ \citenamefont
  {Ueda}}]{Uchino:2010pf}%
  \BibitemOpen
  \bibfield  {author} {\bibinfo {author} {\bibfnamefont {S.}~\bibnamefont
  {Uchino}}, \bibinfo {author} {\bibfnamefont {M.}~\bibnamefont {Kobayashi}},
  \bibinfo {author} {\bibfnamefont {M.}~\bibnamefont {Nitta}}, \ and\ \bibinfo
  {author} {\bibfnamefont {M.}~\bibnamefont {Ueda}},\ }\href {\doibase
  10.1103/PhysRevLett.105.230406} {\bibfield  {journal} {\bibinfo  {journal}
  {Phys. Rev. Lett.}\ }\textbf {\bibinfo {volume} {105}},\ \bibinfo {pages}
  {230406} (\bibinfo {year} {2010})},\ \Eprint {http://arxiv.org/abs/1010.2864}
  {arXiv:1010.2864 [cond-mat.quant-gas]} \BibitemShut {NoStop}%
\bibitem [{\citenamefont {Kobayashi}\ \emph {et~al.}(2012)\citenamefont
  {Kobayashi}, \citenamefont {Kobayashi}, \citenamefont {Kawaguchi},
  \citenamefont {Nitta},\ and\ \citenamefont {Ueda}}]{Kobayashi:2011xb}%
  \BibitemOpen
  \bibfield  {author} {\bibinfo {author} {\bibfnamefont {S.}~\bibnamefont
  {Kobayashi}}, \bibinfo {author} {\bibfnamefont {M.}~\bibnamefont
  {Kobayashi}}, \bibinfo {author} {\bibfnamefont {Y.}~\bibnamefont
  {Kawaguchi}}, \bibinfo {author} {\bibfnamefont {M.}~\bibnamefont {Nitta}}, \
  and\ \bibinfo {author} {\bibfnamefont {M.}~\bibnamefont {Ueda}},\ }\href
  {\doibase 10.1016/j.nuclphysb.2011.11.003} {\bibfield  {journal} {\bibinfo
  {journal} {Nucl. Phys.}\ }\textbf {\bibinfo {volume} {B856}},\ \bibinfo
  {pages} {577} (\bibinfo {year} {2012})},\ \Eprint
  {http://arxiv.org/abs/1110.1478} {arXiv:1110.1478 [math-ph]} \BibitemShut
  {NoStop}%
\bibitem [{\citenamefont {Cipriani}\ \emph {et~al.}(2012)\citenamefont
  {Cipriani}, \citenamefont {Vinci},\ and\ \citenamefont
  {Nitta}}]{Cipriani:2012hr}%
  \BibitemOpen
  \bibfield  {author} {\bibinfo {author} {\bibfnamefont {M.}~\bibnamefont
  {Cipriani}}, \bibinfo {author} {\bibfnamefont {W.}~\bibnamefont {Vinci}}, \
  and\ \bibinfo {author} {\bibfnamefont {M.}~\bibnamefont {Nitta}},\ }\href
  {\doibase 10.1103/PhysRevD.86.121704} {\bibfield  {journal} {\bibinfo
  {journal} {Phys. Rev.}\ }\textbf {\bibinfo {volume} {D86}},\ \bibinfo {pages}
  {121704} (\bibinfo {year} {2012})},\ \Eprint {http://arxiv.org/abs/1208.5704}
  {arXiv:1208.5704 [hep-ph]} \BibitemShut {NoStop}%
\bibitem [{\citenamefont {Alford}\ \emph {et~al.}(2019)\citenamefont {Alford},
  \citenamefont {Baym}, \citenamefont {Fukushima}, \citenamefont {Hatsuda},\
  and\ \citenamefont {Tachibana}}]{Alford:2018mqj}%
  \BibitemOpen
  \bibfield  {author} {\bibinfo {author} {\bibfnamefont {M.~G.}\ \bibnamefont
  {Alford}}, \bibinfo {author} {\bibfnamefont {G.}~\bibnamefont {Baym}},
  \bibinfo {author} {\bibfnamefont {K.}~\bibnamefont {Fukushima}}, \bibinfo
  {author} {\bibfnamefont {T.}~\bibnamefont {Hatsuda}}, \ and\ \bibinfo
  {author} {\bibfnamefont {M.}~\bibnamefont {Tachibana}},\ }\href {\doibase
  10.1103/PhysRevD.99.036004} {\bibfield  {journal} {\bibinfo  {journal} {Phys.
  Rev.}\ }\textbf {\bibinfo {volume} {D99}},\ \bibinfo {pages} {036004}
  (\bibinfo {year} {2019})},\ \Eprint {http://arxiv.org/abs/1803.05115}
  {arXiv:1803.05115 [hep-ph]} \BibitemShut {NoStop}%
\bibitem [{\citenamefont {Chatterjee}\ \emph {et~al.}(2019)\citenamefont
  {Chatterjee}, \citenamefont {Nitta},\ and\ \citenamefont
  {Yasui}}]{Chatterjee:2018nxe}%
  \BibitemOpen
  \bibfield  {author} {\bibinfo {author} {\bibfnamefont {C.}~\bibnamefont
  {Chatterjee}}, \bibinfo {author} {\bibfnamefont {M.}~\bibnamefont {Nitta}}, \
  and\ \bibinfo {author} {\bibfnamefont {S.}~\bibnamefont {Yasui}},\ }\href
  {\doibase 10.1103/PhysRevD.99.034001} {\bibfield  {journal} {\bibinfo
  {journal} {Phys. Rev.}\ }\textbf {\bibinfo {volume} {D99}},\ \bibinfo {pages}
  {034001} (\bibinfo {year} {2019})},\ \Eprint
  {http://arxiv.org/abs/1806.09291} {arXiv:1806.09291 [hep-ph]} \BibitemShut
  {NoStop}%
\bibitem [{\citenamefont {Cherman}\ \emph {et~al.}()\citenamefont {Cherman},
  \citenamefont {Sen},\ and\ \citenamefont {Yaffe}}]{Cherman:2018jir}%
  \BibitemOpen
  \bibfield  {author} {\bibinfo {author} {\bibfnamefont {A.}~\bibnamefont
  {Cherman}}, \bibinfo {author} {\bibfnamefont {S.}~\bibnamefont {Sen}}, \ and\
  \bibinfo {author} {\bibfnamefont {L.~G.}\ \bibnamefont {Yaffe}},\ }\href@noop
  {} {\ }\Eprint {http://arxiv.org/abs/1808.04827} {arXiv:1808.04827 [hep-th]}
  \BibitemShut {NoStop}%
\bibitem [{\citenamefont {Eto}\ \emph {et~al.}(2014)\citenamefont {Eto},
  \citenamefont {Hirono}, \citenamefont {Nitta},\ and\ \citenamefont
  {Yasui}}]{Eto:2013hoa}%
  \BibitemOpen
  \bibfield  {author} {\bibinfo {author} {\bibfnamefont {M.}~\bibnamefont
  {Eto}}, \bibinfo {author} {\bibfnamefont {Y.}~\bibnamefont {Hirono}},
  \bibinfo {author} {\bibfnamefont {M.}~\bibnamefont {Nitta}}, \ and\ \bibinfo
  {author} {\bibfnamefont {S.}~\bibnamefont {Yasui}},\ }\href {\doibase
  10.1093/ptep/ptt095} {\bibfield  {journal} {\bibinfo  {journal} {PTEP}\
  }\textbf {\bibinfo {volume} {2014}},\ \bibinfo {pages} {012D01} (\bibinfo
  {year} {2014})},\ \Eprint {http://arxiv.org/abs/1308.1535} {arXiv:1308.1535
  [hep-ph]} \BibitemShut {NoStop}%
\bibitem [{\citenamefont {Semenoff}\ and\ \citenamefont
  {Zhou}(2007)}]{Semenoff:2006vv}%
  \BibitemOpen
  \bibfield  {author} {\bibinfo {author} {\bibfnamefont {G.~W.}\ \bibnamefont
  {Semenoff}}\ and\ \bibinfo {author} {\bibfnamefont {F.}~\bibnamefont
  {Zhou}},\ }\href {\doibase 10.1103/PhysRevLett.98.100401} {\bibfield
  {journal} {\bibinfo  {journal} {Phys. Rev. Lett.}\ }\textbf {\bibinfo
  {volume} {98}},\ \bibinfo {pages} {100401} (\bibinfo {year} {2007})},\
  \Eprint {http://arxiv.org/abs/cond-mat/0610162} {arXiv:cond-mat/0610162
  [cond-mat]} \BibitemShut {NoStop}%
\bibitem [{\citenamefont {Kobayashi}\ \emph {et~al.}(2009)\citenamefont
  {Kobayashi}, \citenamefont {Kawaguchi}, \citenamefont {Nitta},\ and\
  \citenamefont {Ueda}}]{Kobayashi:2008pk}%
  \BibitemOpen
  \bibfield  {author} {\bibinfo {author} {\bibfnamefont {M.}~\bibnamefont
  {Kobayashi}}, \bibinfo {author} {\bibfnamefont {Y.}~\bibnamefont
  {Kawaguchi}}, \bibinfo {author} {\bibfnamefont {M.}~\bibnamefont {Nitta}}, \
  and\ \bibinfo {author} {\bibfnamefont {M.}~\bibnamefont {Ueda}},\ }\href
  {\doibase 10.1103/PhysRevLett.103.115301} {\bibfield  {journal} {\bibinfo
  {journal} {Phys. Rev. Lett.}\ }\textbf {\bibinfo {volume} {103}},\ \bibinfo
  {pages} {115301} (\bibinfo {year} {2009})},\ \Eprint
  {http://arxiv.org/abs/0810.5441} {arXiv:0810.5441 [cond-mat.other]}
  \BibitemShut {NoStop}%
\bibitem [{\citenamefont {Mizushima}\ and\ \citenamefont
  {Nitta}(2018)}]{Mizushima:2017pma}%
  \BibitemOpen
  \bibfield  {author} {\bibinfo {author} {\bibfnamefont {T.}~\bibnamefont
  {Mizushima}}\ and\ \bibinfo {author} {\bibfnamefont {M.}~\bibnamefont
  {Nitta}},\ }\href {\doibase 10.1103/PhysRevB.97.024506} {\bibfield  {journal}
  {\bibinfo  {journal} {Phys. Rev.}\ }\textbf {\bibinfo {volume} {B97}},\
  \bibinfo {pages} {024506} (\bibinfo {year} {2018})},\ \Eprint
  {http://arxiv.org/abs/1710.07403} {arXiv:1710.07403 [cond-mat.supr-con]}
  \BibitemShut {NoStop}%
\bibitem [{\citenamefont {Venderbos}\ \emph {et~al.}(2018)\citenamefont
  {Venderbos}, \citenamefont {Savary}, \citenamefont {Ruhman}, \citenamefont
  {Lee},\ and\ \citenamefont {Fu}}]{PhysRevX.8.011029}%
  \BibitemOpen
  \bibfield  {author} {\bibinfo {author} {\bibfnamefont {J.~W.~F.}\
  \bibnamefont {Venderbos}}, \bibinfo {author} {\bibfnamefont {L.}~\bibnamefont
  {Savary}}, \bibinfo {author} {\bibfnamefont {J.}~\bibnamefont {Ruhman}},
  \bibinfo {author} {\bibfnamefont {P.~A.}\ \bibnamefont {Lee}}, \ and\
  \bibinfo {author} {\bibfnamefont {L.}~\bibnamefont {Fu}},\ }\href {\doibase
  10.1103/PhysRevX.8.011029} {\bibfield  {journal} {\bibinfo  {journal} {Phys.
  Rev. X}\ }\textbf {\bibinfo {volume} {8}},\ \bibinfo {pages} {011029}
  (\bibinfo {year} {2018})}\BibitemShut {NoStop}%
\end{thebibliography}%

\end{document}